\pdfoutput=1

\documentclass[11pt,twoside,a4paper,cmspaper,final,collab]{cms-tdr}
\def\svnVersion{455001P}\def\cmsCernNoTag{CERN-EP-2017-238}\def\cmsCernDate{\today}\def\cmsMessage{Published in Physics Letters B as \href{http://dx.doi.org/10.1016/j.physletb.2018.03.084}{\doi{10.1016/j.physletb.2018.03.084.}}}

\begin{document}\cmsNoteHeader{B2G-16-026}

\hyphenation{had-ron-i-za-tion}
\hyphenation{cal-or-i-me-ter}
\hyphenation{de-vices}
\RCS$Revision: 452452 $
\RCS$HeadURL: svn+ssh://svn.cern.ch/reps/tdr2/papers/B2G-16-026/trunk/B2G-16-026.tex $
\RCS$Id: B2G-16-026.tex 452452 2018-03-23 21:36:33Z alverson $
\newlength\cmsFigWidth
\ifthenelse{\boolean{cms@external}}{\setlength\cmsFigWidth{0.85\columnwidth}}{\setlength\cmsFigWidth{0.4\textwidth}}
\ifthenelse{\boolean{cms@external}}{\providecommand{\cmsLeft}{upper\xspace}}{\providecommand{\cmsLeft}{left\xspace}}
\ifthenelse{\boolean{cms@external}}{\providecommand{\cmsRight}{lower\xspace}}{\providecommand{\cmsRight}{right\xspace}}
\ifthenelse{\boolean{cms@external}}{\providecommand{\cmsLeftU}{upper\xspace}}{\providecommand{\cmsLeftU}{upper left\xspace}}
\ifthenelse{\boolean{cms@external}}{\providecommand{\cmsRightU}{middle\xspace}}{\providecommand{\cmsRightU}{upper right\xspace}}

\newcommand{\Hbb}{\ensuremath{\PH\to\bbbar}\xspace}
\newcommand{\ttjets}{\ttbar+jets\xspace}
\newcommand{\intLumi}{35.9\fbinv}
\newcommand{\Hbbt}{double-\PQb\,tagger\xspace}
\newcommand{\mjj}{\ensuremath{m_{\text{jj}}}\xspace}
\newcommand{\mjjs}{\ensuremath{m_{\text{jj,red}}}\xspace}
\newcommand{\mjjred}{\ensuremath{m_{\text{jj,red}}}\xspace}
\newcommand{\mH}{\ensuremath{m_{\PH}}\xspace}
\newcommand{\nsub}{\ensuremath{\tau_{21}}\xspace}
\newcommand{\mjone}{\ensuremath{m_{\text{j}_{1}}}\xspace}
\newcommand{\mjtwo}{\ensuremath{m_{\text{j}_{2}}}\xspace}
\newcommand{\LambdaR}{\ensuremath{\Lambda_{\text{R}}}\xspace}
\newcommand{\DeltaEta}{\ensuremath{\Delta\eta(\text{j}_{1}, \text{j}_{2})}\xspace}
\newcommand{\PX}{\ensuremath{\cmsSymbolFace{X}}}
\newcommand{\mx}{\ensuremath{m_{\PX}}\xspace}
\newcolumntype{P}[1]{>{\centering\arraybackslash}p{#1}}

\cmsNoteHeader{B2G-16-026}
\title{Search for a massive resonance decaying to a pair of Higgs bosons in the four b quark final state in proton-proton collisions at $\sqrt{s}=13$\TeV}

\date{\today}

\abstract{
A search for a massive resonance decaying into a pair of standard model Higgs bosons,
in a final state consisting of two b quark-antiquark pairs, is performed.
A data sample of proton-proton collisions at a centre-of-mass energy of 13\TeV is used,
collected by the CMS experiment at the CERN LHC in 2016,
and corresponding to an integrated luminosity of 35.9\fbinv.
The Higgs bosons are highly Lorentz-boosted and are each reconstructed as a single large-area jet.
The signal is characterized by a peak in the dijet invariant mass distribution,
above a background from the standard model multijet production.
The observations are consistent with the background expectations,
and are interpreted as upper limits on the products of the $s$-channel production
cross sections and branching fractions of narrow bulk gravitons and
 radions in warped extra-dimensional models. The limits range from
126 to 1.4\unit{fb} at 95\% confidence level for resonances with masses between 750 and 3000\GeV,
and are the most stringent to date, over the explored mass range.}

\hypersetup{%
pdfauthor={CMS Collaboration},%
pdftitle={Search for a massive resonance decaying to a pair of Higgs bosons in the four b quark final state in proton-proton collisions at sqrt(s)=13 TeV},%
pdfsubject={CMS},%
pdfkeywords={CMS, physics, extradimensions, graviton, radion, di-Higgs boson resonance}}

\maketitle

\section{Introduction\label{sec:Introduction}}

In the standard model (SM), the pair production of Higgs bosons ($\PH$)~\cite{HiggsDiscoveryAtlas,HiggsDiscoveryCMS,CMSHiggsLongPaper} in proton-proton ($\Pp\Pp$) collisions at $\sqrt{s} = 13\TeV$ is a rare process~\cite{deFlorian:2013jea}.
However, the existence of massive resonances decaying to Higgs boson pairs ($\PH\PH$) in many new physics models may enhance this rate to a level observable at the CERN LHC using the current data.
For instance, models with warped extra dimensions (WED)~\cite{Randall:1999ee} contain new particles such as the spin-0 radion~\cite{Goldberger:1999uk,Csaki:1999mp,Csaki:2000zn} and the spin-2 first Kaluza--Klein (KK) excitation of the graviton~\cite{Davoudiasl:1999jd,DeWolfe:1999cp, Agashe:2007zd}, which have sizeable branching fractions to $\PH\PH$.

The WED models have an extra spatial dimension compactified between two branes, with the region between (called the bulk) warped via an exponential metric ${\kappa l}$, $\kappa$ being the warp factor and $l$ the coordinate of the extra spatial dimension~\cite{Giudice:2000av}. The reduced Planck scale ($\overline{\Mpl} \equiv \Mpl/8\pi$, \Mpl being the Planck scale) is considered a fundamental scale.
The free parameters of the model are $\kappa/\overline{\Mpl}$ and the ultraviolet cutoff of the theory $\LambdaR \equiv \sqrt{6} \re^{-\kappa l} \overline{\Mpl}$~\cite{Goldberger:1999uk}.
In $\Pp\Pp$ collisions at the LHC, the graviton and the radion are produced primarily through gluon-gluon fusion and are predicted to decay to $\PH\PH$~\cite{Oliveira:2014kla}.

Other scenarios, such as the two-Higgs doublet models~\cite{Branco:2011iw} (in particular, the minimal supersymmetric model~\cite{Djouadi:2005gj}) and the Georgi-Machacek model~\cite{GEORGI1985463} predict spin-0 resonances that are produced primarily through gluon-gluon fusion, and decay to an $\PH\PH$ pair.
These particles have the same Lorentz structure and effective couplings to the gluons and, for narrow widths, result in the same kinematic distributions as those for the bulk radion. Hence, the results of this paper are also applicable to this class of models.

Searches for a new particle $\PX$ in the $\PH\PH$ decay channel have been performed by the
ATLAS~\cite{Aad:2014yja, Aad:2015uka, Aad:2015xja} and
CMS~\cite{Khachatryan:2014jya, Khachatryan:2015year, Khachatryan:2015tha,Khachatryan:2016sey,Khachatryan:2016cfa}
Collaborations in $\Pp\Pp$ collisions at $\sqrt{s} =  7$ and 8\TeV.
More recently, the ATLAS Collaboration has published limits on the production of a KK bulk graviton, decaying to $\PH\PH$, in the $\bbbar\bbbar$ final state, using $\Pp\Pp$ collision data at $\sqrt{s} = 13\TeV$, corresponding to an integrated luminosity of 3.2\fbinv~\cite{Aaboud:2016xco}. Because the longitudinal components of the $\PW$ and $\PZ$ bosons couple to the Higgs field in the SM, a resonance decaying to $\PH\PH$ potentially also decays into $\PW\PW$ and $\PZ\PZ$, with a comparable branching fraction for $\PX\to\PZ\PZ$, and with a branching fraction for $\PX\to\PW\PW$ that is twice as large.
Searches for $\PX \to \PW\PW$ and $\PZ\PZ$ have been performed by ATLAS and CMS~\cite{ATLASVV,ATLASWV,ATLASZV,Khachatryan:2014hpa,CMSZVWV,ATLAS13TeV_WW_WZ_ZZ_allhad,ATLAS13TeV_WW_WZ_semilep,ATLAS13TeV_ZZ_WZ_dilep_met,CMSVV_8_13combo,CMS_13TeV_WW_WZ_ZZ_qW_qZ_allhad}.

This letter reports on the search for a massive resonance decaying to an $\PH\PH$ pair, in the $\bbbar\bbbar$ final state (with a branching fraction $\approx$33\%~\cite{deFlorian:2016spz}), performed using a data set corresponding to \intLumi of $\Pp\Pp$ collisions at $\sqrt{s} = 13\TeV$.
The search significantly improves upon the CMS analysis performed using the LHC data collected at $\sqrt{s} = 8\TeV$~\cite{Khachatryan:2016cfa}, and extends the searched mass range to 750--3000\GeV.
This search is conducted for both the radion and the graviton, whereas the earlier search only considered the former.

In this search, the $\PX \to \PH\PH$ decay would result in highly Lorentz-boosted and collimated decay products of \Hbb, which are referred to as $\PH$ jets. These are reconstructed using jet substructure and jet flavour-tagging techniques~\cite{Butterworth:2008iy,Cooper:2013kia,Gouzevitch:2013qca}.
The background consists mostly of SM multijet events, and is estimated using several control regions defined in the phase space of the masses and flavour-tagging discriminators of the two $\PH$ jets, and the $\PH\PH$ dijet invariant mass, allowing the background to be predicted over the entire range of $\mx$ explored.
The signal would appear as a peak in the $\PH\PH$ dijet invariant mass spectrum above a smooth background distribution.
\section{The CMS detector and event simulations\label{sec:CMSDetector}}

The CMS detector with its coordinate system and the relevant kinematic
variables is described in Ref.~\cite{CMSDetector}. The central
feature of the CMS apparatus is a superconducting solenoid of
6\unit{m} internal diameter, providing a magnetic field of
3.8\unit{T}. Within the field volume are silicon pixel and strip
trackers, a lead tungstate crystal electromagnetic calorimeter (ECAL),
and a brass and scintillator hadron calorimeter (HCAL), each composed
of a barrel and two endcap sections. The tracker covers a
pseudorapidity $\eta$ range from $-2.5$ to 2.5 with the ECAL and the
HCAL extending up to $\abs{\eta} = 3$. Forward calorimeters in the region
up to $\abs{\eta} = 5$ provide almost hermetic detector coverage.
Muons are detected in gas-ionization chambers embedded
in the steel flux-return yoke outside the solenoid, covering a region of
$\abs{\eta} < 2.4$.

Events of interest are selected using a two-tiered trigger system~\cite{CMSTrigger}. The first level (L1), composed of custom hardware processors, uses information from the calorimeters and muon detectors to select events at a rate of around 100\unit{kHz}. The second level, known as the high-level trigger (HLT), consists of a farm of processors running a version of the full event reconstruction software optimized for fast processing, and reduces the event rate to around 1\unit{kHz} before data storage. Events are selected at the trigger level by the presence of jets of particles in the detector. The L1 trigger algorithms reconstruct jets from energy deposits in the calorimeters. At the HLT, physics objects (charged and neutral hadrons, electrons, muons, and photons) are reconstructed using a particle-flow (PF) algorithm~\cite{CMS-PF-GED}. The anti-\kt algorithm~\cite{antiKtAlgorithm, FastJet} is used to cluster these objects with a distance parameter of 0.8 (AK8 jets) or 0.4 (AK4 jets).

{\tolerance=2400
Bulk graviton and radion signal events are simulated at leading order using the {\MGvATNLO 2.3.3}~\cite{MG5_aMCNLO} event generator for masses
in the range 750--3000\GeV and widths of 1\MeV (narrow width approximation).
The {NNPDF3.0} leading order parton distribution functions (PDFs)~\cite{Ball:2014uwa}, taken from the
{LHAPDF6} PDF set~\cite{Harland-Lang:2014zoa,Buckley:2014ana,Carrazza:2015hva,Butterworth:2015oua}, with the four-flavour scheme, is used. The showering and hadronization of partons is simulated with
\PYTHIA~8.212~\cite{Pythia8p2}.
The \HERWIGpp~2.7.1~\cite{Bahr:2008pv} generator is used for an
alternative model to evaluate the systematic uncertainty
associated with the parton shower and hadronization.
The tune {CUETP8M1-NNPDF2.3LO}~\cite{CUETTune} is used for \PYTHIA~8,
while the EE5C tune~\cite{HerwigppEE5C} is used for \HERWIGpp.
\par}

The background is modelled entirely from data.
However, simulated background samples are used to develop and validate the background estimation techniques, prior to being applied to the data.
These are multijet events, generated at leading order using \MGvATNLO,
and  \ttjets, generated at next-to-leading order using {\POWHEG~2.0}~\cite{POWHEG, POWHEG_Frixione, POWHEGBOX}.
Both these backgrounds are interfaced to \PYTHIA~8 for simulating the parton shower and hadronization.
Studies using simulations established that the multijet component is more than 99\% of the background, with the rest mostly from \ttjets production.

All generated samples were processed through a \GEANTfour-based~\cite{Agostinelli:2002hh,GEANT} simulation of the CMS
detector. Multiple $\Pp\Pp$ collisions may occur in the same or adjacent LHC bunch crossings
(pileup) and contribute to the overall event activity in the
detector. This effect is included in the simulations,
and the samples are reweighted to match the number of $\Pp\Pp$
interactions observed in the data, assuming a total inelastic
$\Pp\Pp$ collision cross section of 69.2\unit{mb}~\cite{CMSLumi13TeV}.

\section{Event selection\label{sec:EvtSel}}

Events were collected using several HLT algorithms.
The first required the scalar $\pt$ sum of all AK4 jets in the event ($\HT$) to be greater than 800 or 900\GeV, depending on the LHC beam instantaneous luminosity.
A second trigger criterion required $\HT \ge 650\GeV$, with a pair of AK4 jets with invariant mass above 900\GeV and a pseudorapidity separation $|\Delta\eta| < 1.5$.
A third set of triggers selected events with the scalar $\pt$ sum of all AK8 jets greater than 650 or 700\GeV and the presence of an AK8 jet with a ``trimmed mass'' above 50\GeV, \ie the jet mass after removing remnants of soft radiation using jet trimming technique~\cite{Krohn:2009th}.
The fourth triggering condition was based on the presence of an AK8 jet with $\pt > 360\GeV$ and trimmed mass greater than 30\GeV.
The last trigger selection accepted events containing two AK8 jets having $\pt > 280$ and $200\GeV$ with at least one having trimmed mass greater than 30\GeV, together with an AK4 jet passing a loose $\PQb$-tagging criterion.

The $\Pp\Pp$ interaction vertex with the highest $\sum\pt^{2}$ of the associated clusters of physics objects is considered to be the one associated with the hard scattering interaction, the primary vertex.
The physics objects are the jets, clustered using the jet finding algorithm~\cite{antiKtAlgorithm,FastJet} with the tracks assigned to the vertex as inputs, and the associated missing transverse momentum, taken as the negative vector sum of the $\pt$ of those jets.
The other interaction vertices are designated as pileup vertices.

To mitigate the effect of pileup, particles are assigned weights using the pileup per
particle identification (PUPPI) algorithm~\cite{PUPPI}, with the
weight corresponding to its estimated probability to originate from a pileup
interaction. Charged particles from pileup vertices receive a weight of zero while
those from the primary vertex receive a weight of one. Neutral particles
are assigned a weight between zero and one, with higher values for those
likely to originate from the primary vertex. Particles are
then clustered into AK8 jets.
The vector sum of the weighted momenta of
all particles clustered in the jet is taken to be the jet momentum. To
account for detector response nonlinearity, jet energy corrections are applied
as a function of jet $\eta$ and $\pt$~\cite{Khachatryan:2016kdb,CMS-DP-2016-020}.
In each event, the leading and the subleading $\pt$ AK8 jets, j$_{1}$ and j$_{2}$, respectively, are required to have $\pt > 300\GeV$ and $\abs{\eta} < 2.4$.

The removal of events containing isolated leptons (electrons or muons) with $\pt > 20\GeV$ and $\abs{\eta} < 2.4$ helps suppress \ttjets and diboson backgrounds.
The isolation variable is defined as the scalar $\pt$ sum of the charged and neutral hadrons, and photons in a cone of $\Delta R = 0.3$ for an electron or $\Delta R = 0.4$ for a muon, where $\Delta R \equiv \sqrt{\smash[b]{(\Delta\eta)^2 + (\Delta\phi)^2}}$, $\phi$ being the azimuthal angle in radians. The energy from pileup deposited in the isolation cone, and the $\pt$ of the lepton itself, is subtracted~\cite{CMSElectronReco, CMSMuonReco}.
The isolation requirement removes jets misidentified as leptons.
Additional quality criteria are applied to improve the purity of the isolated lepton samples.
Electrons passing combined isolation and quality criteria corresponding to a selection efficiency of 90\% (70\%) are designated  ``loose'' (``medium'') electrons. For the ``loose'' (``medium'') muons, the total associated efficiency is 100\% (95\%). The probability of a jet to be misidentified as an electron or a muon is in the range 0.5--2\%, depending on $\pt$, $\eta$, and the choice of medium or loose selection criteria. Events containing one medium lepton, or two loose leptons of the same flavour, but of opposite charge, are rejected.

The $\PH \to \bbbar$ system is reconstructed as a single high-$\pt$
AK8 jet, where the decay products have merged within the jet, and the two
highest $\pt$ jets in the event are assumed to be the Higgs boson
candidates. The jet is groomed~\cite{Salam:2009jx} to remove soft and
wide-angle radiation using the soft-drop algorithm~\cite{Dasgupta:2013ihk,Larkoski:2014wba}, with the soft radiation fraction parameter $z$ set to $0.1$ and the angular
exponent parameter $\beta$ set to 0. The groomed jet is used
to compute the soft-drop jet mass, which peaks at the Higgs boson mass
for signal events and reduces the mass of background quark- and
gluon-initiated jets. Dedicated mass
corrections~\cite{CMS-PAS-JME-16-003}, derived from simulation and
data in a region enriched with $\ttbar$ events with merged
$\PW \to \qqbar$ decays, are applied to the jet mass in order to remove
residual dependence on the jet $\pt$, and to match the jet mass scale
and resolution observed in data.

\begin{figure}[htbp]
\centering
\includegraphics[width=0.45\textwidth]{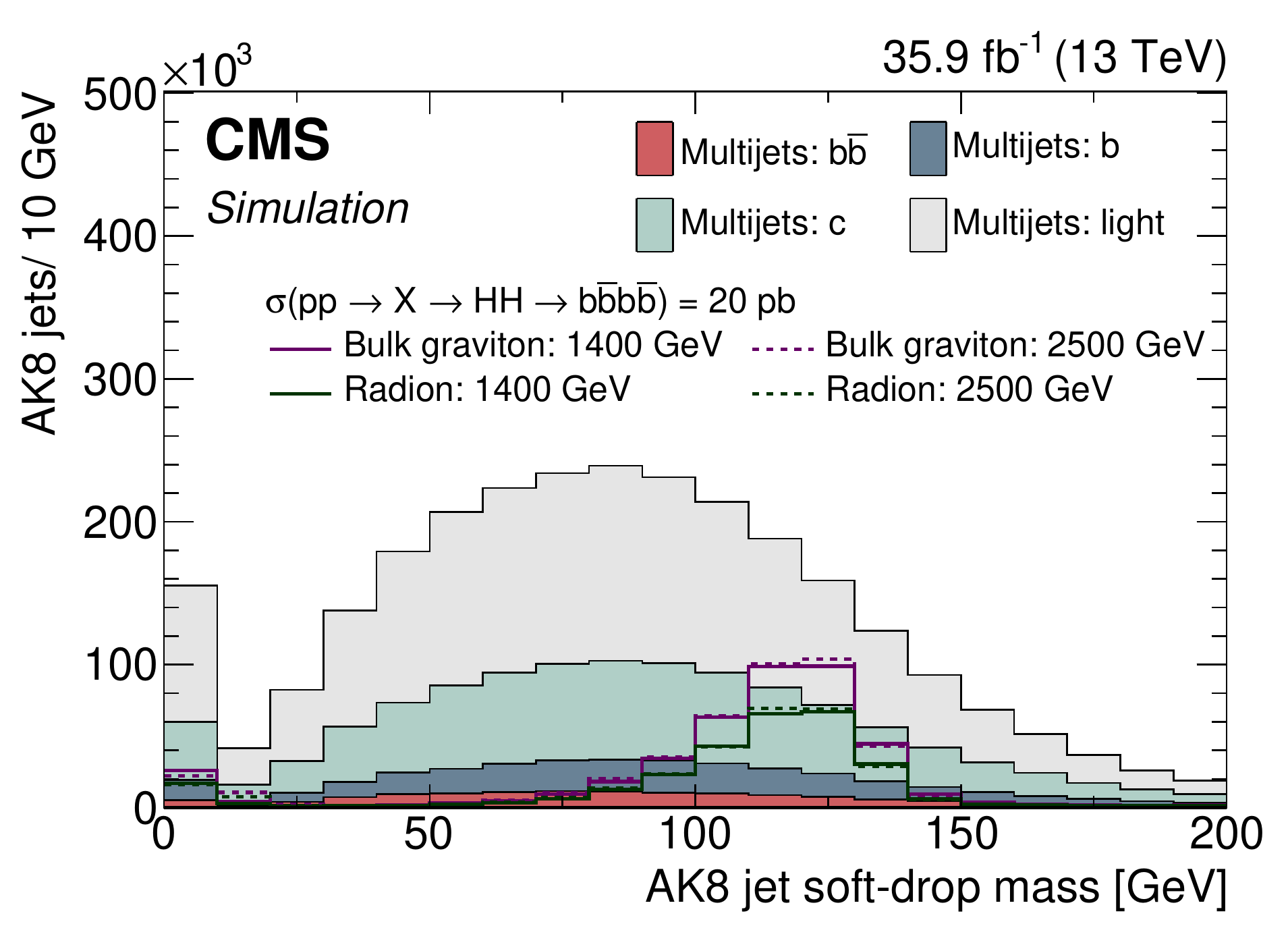}
\includegraphics[width=0.45\textwidth]{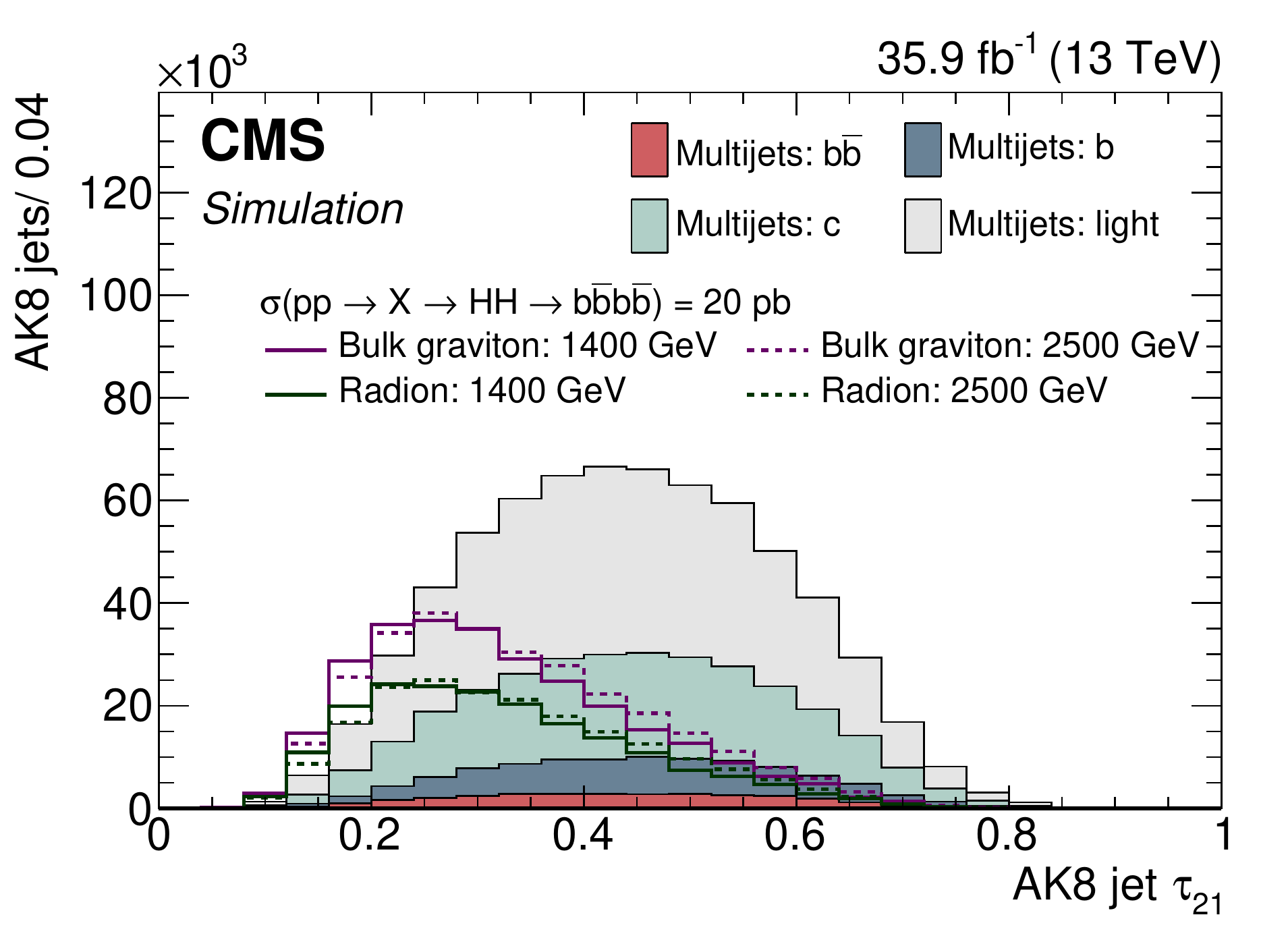}
\includegraphics[width=0.45\textwidth]{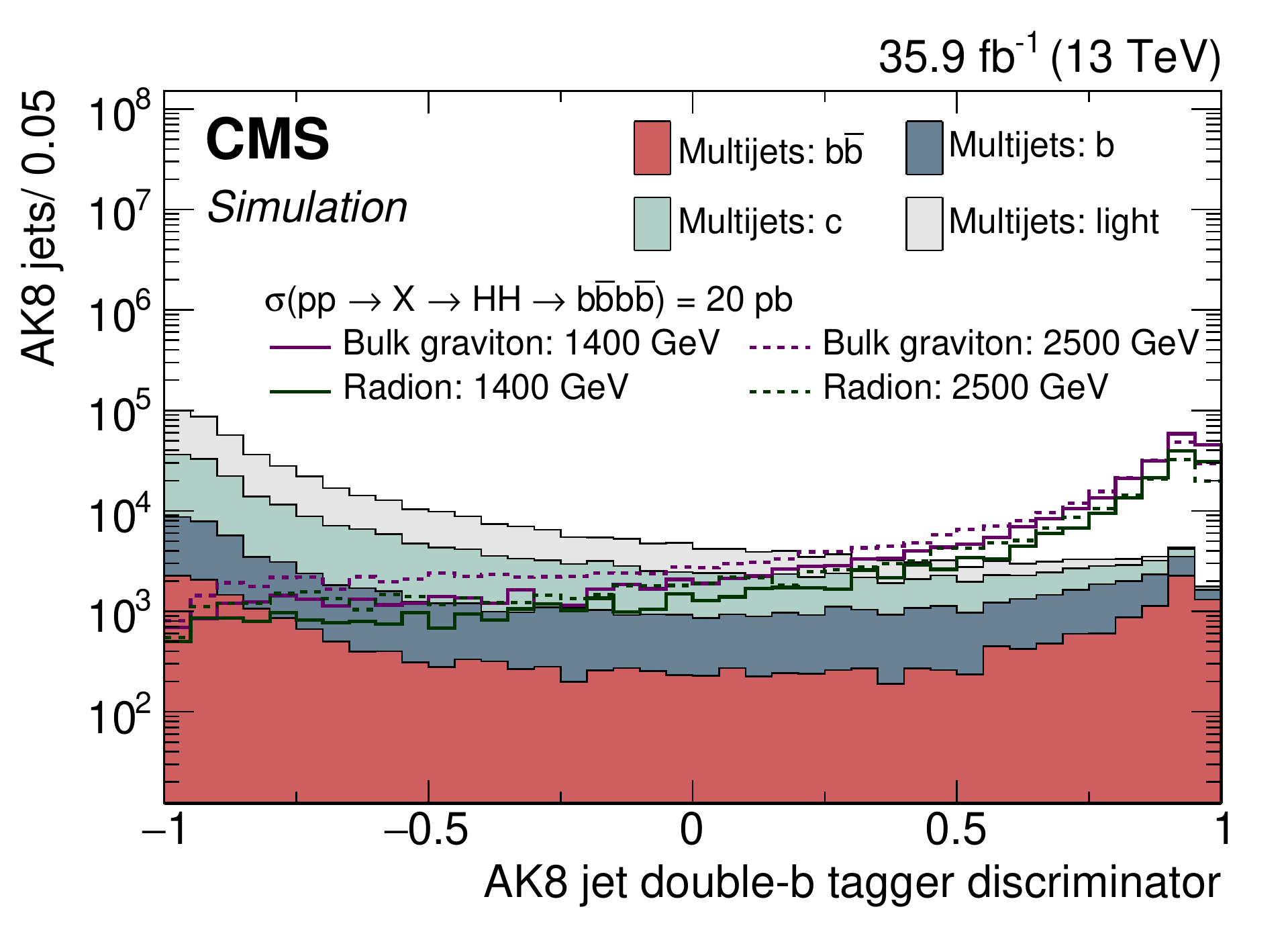}
\caption{The soft-drop mass (\cmsLeftU), the N-subjettiness $\nsub$
  (\cmsRightU), and the \Hbbt discriminator (lower) distributions of the selected AK8
  jets. The multijet background components for the different jet
  flavours are shown: jets having two B hadrons ($\bbbar$) or a single one ($\cPqb$), jets having a charm hadron ($\cPqc$), and all other jets (light).
  Also plotted are the distributions for the simulated bulk graviton and radion signals of
  masses 1400 and 2500\GeV. The number of signal and background
  events correspond to an integrated luminosity of \intLumi. A signal
  cross section $\sigma(\Pp\Pp \to X \to \PH\PH \to \bbbar\bbbar) = 20\unit{pb}$ is assumed for all the mass
  hypotheses. The events are required to have passed the trigger
  selection, lepton rejection, the AK8 jet kinematic selections $\pt >
  300\GeV$ and $\abs{\eta} < 2.4$, and $\abs{\DeltaEta} < 1.3$. The reduced dijet
  invariant mass $\mjjs$ is required to be greater than 750\GeV. The
  N-subjettiness requirement of $\nsub < 0.55$ is applied to the \cmsLeftU and lower figures. The soft-drop masses of the two jets
  are between 105--135\GeV for the \cmsRightU and lower
  figures.\label{fig:uncorrsd_prebtag}}
\end{figure}

\begin{figure*}[htbp]
\centering
\includegraphics[width=0.49\textwidth]{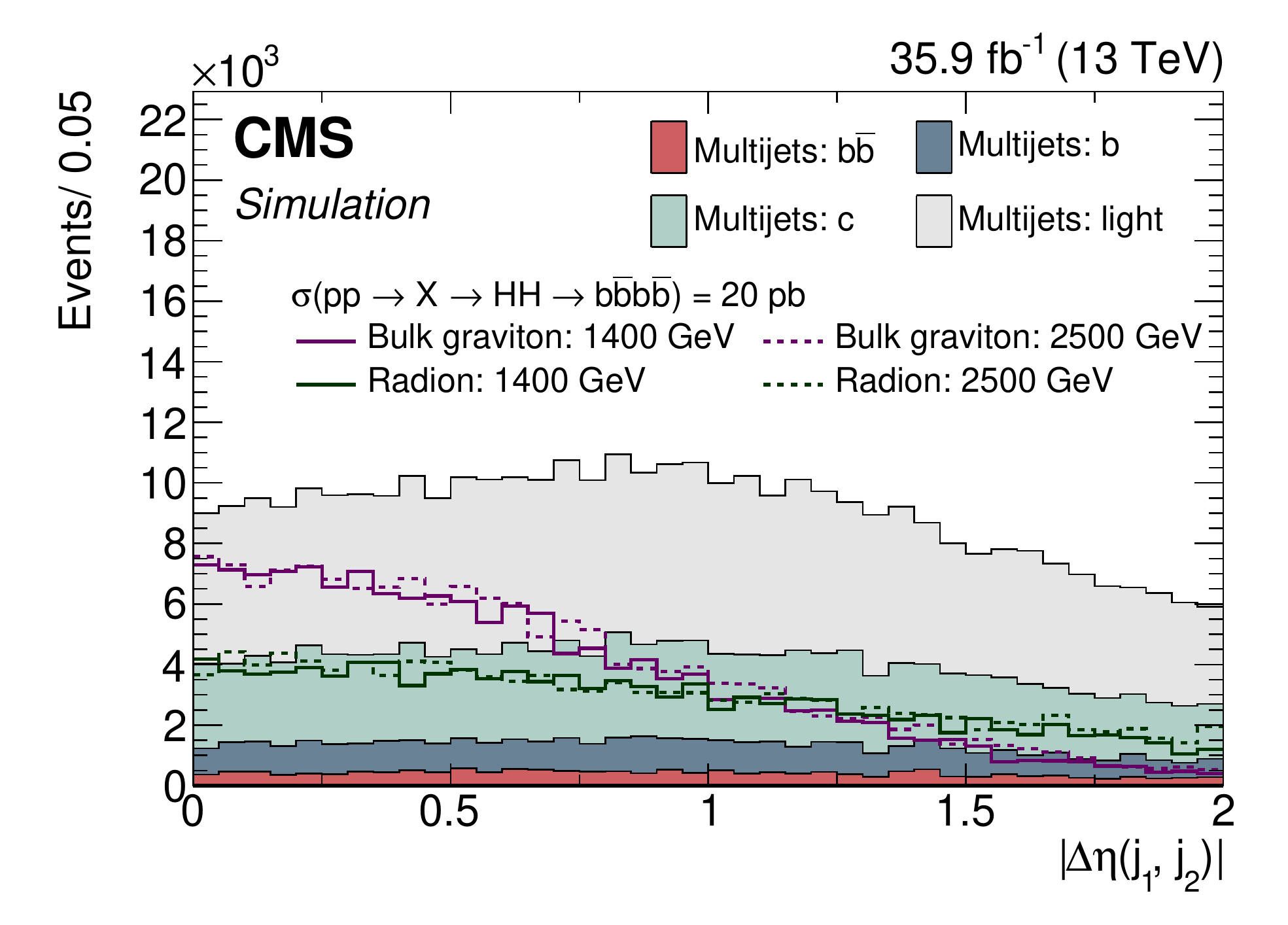}
\includegraphics[width=0.49\textwidth]{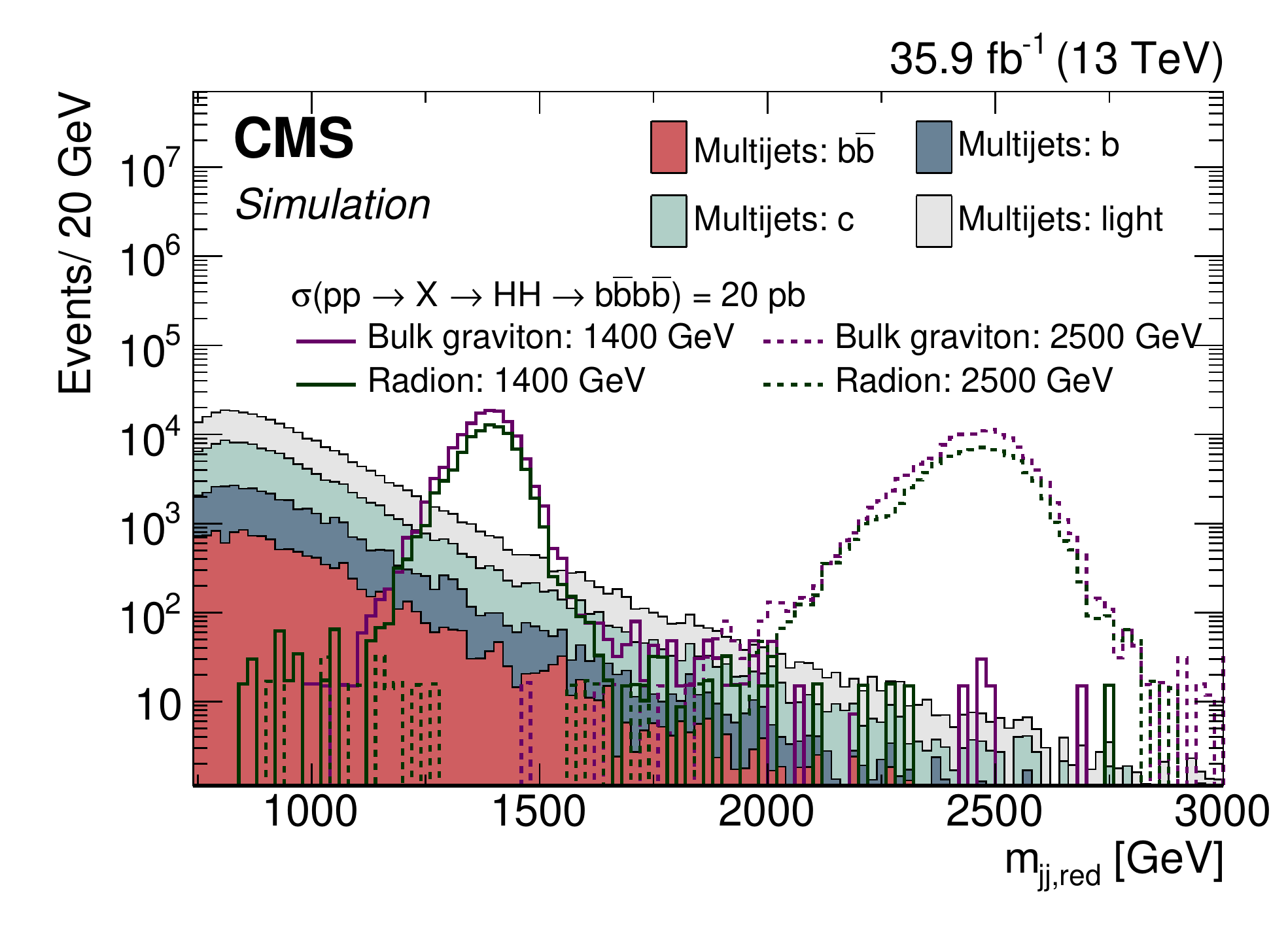}
\caption{The jet separation $\abs{\DeltaEta}$ (left) and the reduced dijet
  invariant mass $\mjjs$ (right) distributions.
  The multijet background components for the different jet flavours are shown:
  events containing at least one jet with two B hadrons ($\bbbar$) or a single one ($\cPqb$), events containing a jet having a charm hadron ($\cPqc$), and all other events (light).
  Also plotted are the distributions for the simulated bulk graviton and radion signals
  of masses 1400 and 2500\GeV. The numbers of signal
  and background events correspond to an integrated luminosity of \intLumi.
  The signal cross section $\sigma(\Pp\Pp \to X \to \PH\PH \to \bbbar\bbbar)$ is assumed to be 20\unit{pb}
  for all the mass hypotheses. The events are required to have passed
  the online selection, lepton rejection, the AK8 jet kinematic
  selections $\pt > 300\GeV$, $\abs{\eta} < 2.4$. The soft-drop masses of
  the two jets are between 105 and 135\GeV, and the N-subjettiness
  requirement of $\nsub < 0.55$ and $\mjjs > 750\GeV$ are applied.
  The $\mjjs$ distributions (right) require $\abs{\DeltaEta} < 1.3$.}
\label{fig:deta_mjjred_prebtag}
\end{figure*}

The soft-drop masses of j$_{1}$ and j$_{2}$ are required to be within the range 105--135\GeV, with an efficiency of about 60--70\%, for jets arising from a signal of mass $\mx$ in the range 750--3000\GeV.
The ``N-subjettiness'' algorithm is used to
determine the consistency of the jet with two subjets from a two-pronged \Hbb decay, by computing the inclusive jet shape variables $\tau_{1}$ and $\tau_{2}$~\cite{Thaler:2011gf}. The ratio $\nsub \equiv \tau_2/\tau_1$ with a value much less than one indicates a jet with two subjets. The selection $\nsub < 0.55$ is used, having a jet $\pt$-dependent efficiency of 50--70\%, before applying the soft-drop mass requirement.

For background events, j$_{1}$ and j$_{2}$ are often well separated in $\eta$, especially at high invariant mass (\mjj) of j$_{1}$ and j$_{2}$.
In contrast, signal events that contain a heavy resonance decaying to two energetic $\PH$ jets are characterized by a small separation of the two jets in $\eta$. Events are therefore required to have a pseudorapidity separation $\abs{\DeltaEta} < 1.3$.

The efficiency of the trigger combination is measured in a sample of multijet events, collected with a control trigger requiring a single AK4 jet with $\pt > 260\GeV$, and with the leading and the subleading $\pt$ AK8 jets, j$_{1}$ and j$_{2}$, respectively, passing the above selections on $\pt$, $\eta$, and the soft-drop mass. The efficiency is greater than 99\% for $\mjj \ge 1100\GeV$, and in the range 40--99\% for $750 < \mjj < 1100\GeV$.
The trigger efficiency of the simulated samples is corrected using a scale factor to match the observed efficiency in the data. This scale factor is applied as a function of $\abs{\DeltaEta}$ because it has a mild dependence on this variable.

The main method to suppress the multijet background is
$\PQb$ tagging: since a true $\PH \to \bbbar$ jet contains two $\PQb$ hadrons,
the $\PH$ jet candidates are identified using the dedicated ``\Hbbt''
algorithm~\cite{CMS-BTV-16-001}. The \Hbbt exploits the presence
of two hadronized $\PQb$ quarks inside the \PH jet, and uses
variables related to $\PQb$ hadron lifetime and mass to distinguish between
\PH jets and the background from multijet production; it also exploits
the fact that the $\PQb$ hadron flight directions are strongly correlated
with the axes used to calculate the N-subjettiness observables.
The \Hbbt is a multivariate discriminator with output between $-1$ and 1,
with a higher value indicating a greater probability for the jet to contain a $\bbbar$ pair.
The \Hbbt discriminator thresholds of 0.3 and 0.8 correspond to $\PH$ jet
tagging efficiencies of 80 and 30\% and are referred to as ``loose''
(L) and ``tight'' (T) requirements, respectively.
Events must have the two leading $\pt$ AK8 jets satisfying the loose \Hbbt requirement.
The data-to-simulation scale factor for the \Hbbt efficiency is measured in an event sample enriched in $\bbbar$ pairs from gluon splitting~\cite{CMS-BTV-16-001}, and applied to the signals to obtain the correct signal yields.

The main variable used in the search for a $\PH\PH$ resonance is the ``reduced dijet invariant mass'' $\mjjs \equiv \mjj - (\mjone - \mH) - (\mjtwo - \mH)$, where \mjone and \mjtwo are the soft-drop masses of the leading and subleading $\PH$-tagged jets in the event, and $\mH = 125.09\GeV$~\cite{Aad:2015zhl,Sirunyan:2017exp} is the Higgs boson mass. The quantity \mjjs is used rather than \mjj since by subtracting the soft-drop masses of the two $\PH$-tagged jets and adding back the exact Higgs boson mass $\mH$, fluctuations in $\mjone$ and $\mjtwo$ due to the jet mass resolution are corrected, leading to 8--10\% improvement in the dijet mass resolution. A requirement of $\mjjs > 750\GeV$ is applied for selecting signal-like events.

The soft-drop mass, $\nsub$, and \Hbbt discriminator
distributions of the two leading $\pt$ jets are shown in
Fig.~\ref{fig:uncorrsd_prebtag} for simulated events after passing the
online selection, lepton rejection, kinematic selection, and the
requirement $\mjjs > 750\GeV$.
Also, the N-subjettiness requirement of $\nsub
< 0.55$ is applied for the soft-drop mass and the \Hbbt
distributions,
while the soft-drop mass requirement is applied to the $\nsub$, and \Hbbt discriminator distributions.
Since some of the triggers impose a trimmed jet mass requirement, this affects the shape of the offline soft-drop jet mass, resulting in a steep rise above $\sim$20\GeV.
The distributions of the $\abs{\DeltaEta}$ and the
\mjjred variables are shown in Fig.~\ref{fig:deta_mjjred_prebtag}.
In these figures, the multijet background is shown for different jet flavour categories: jets having two B hadrons ($\bbbar$) or a single one ($\cPqb$), jets having a charm hadron ($\cPqc$), and all other jets (light).

The \Hbbt discriminator of the two leading AK8 jets must exceed the
loose threshold. In addition, if both discriminator values also exceed the tight threshold, events are classified in the ``TT'' category. Otherwise, they are classified in the ``LL'' category, which contains events with both j$_{1}$ and j$_{2}$ failing the tight threshold as well as events with either j$_{1}$ or j$_{2}$ passing the tight threshold while the other passes the loose threshold only.

\begin{figure}[htb]
\centering
\includegraphics[width=0.49\textwidth]{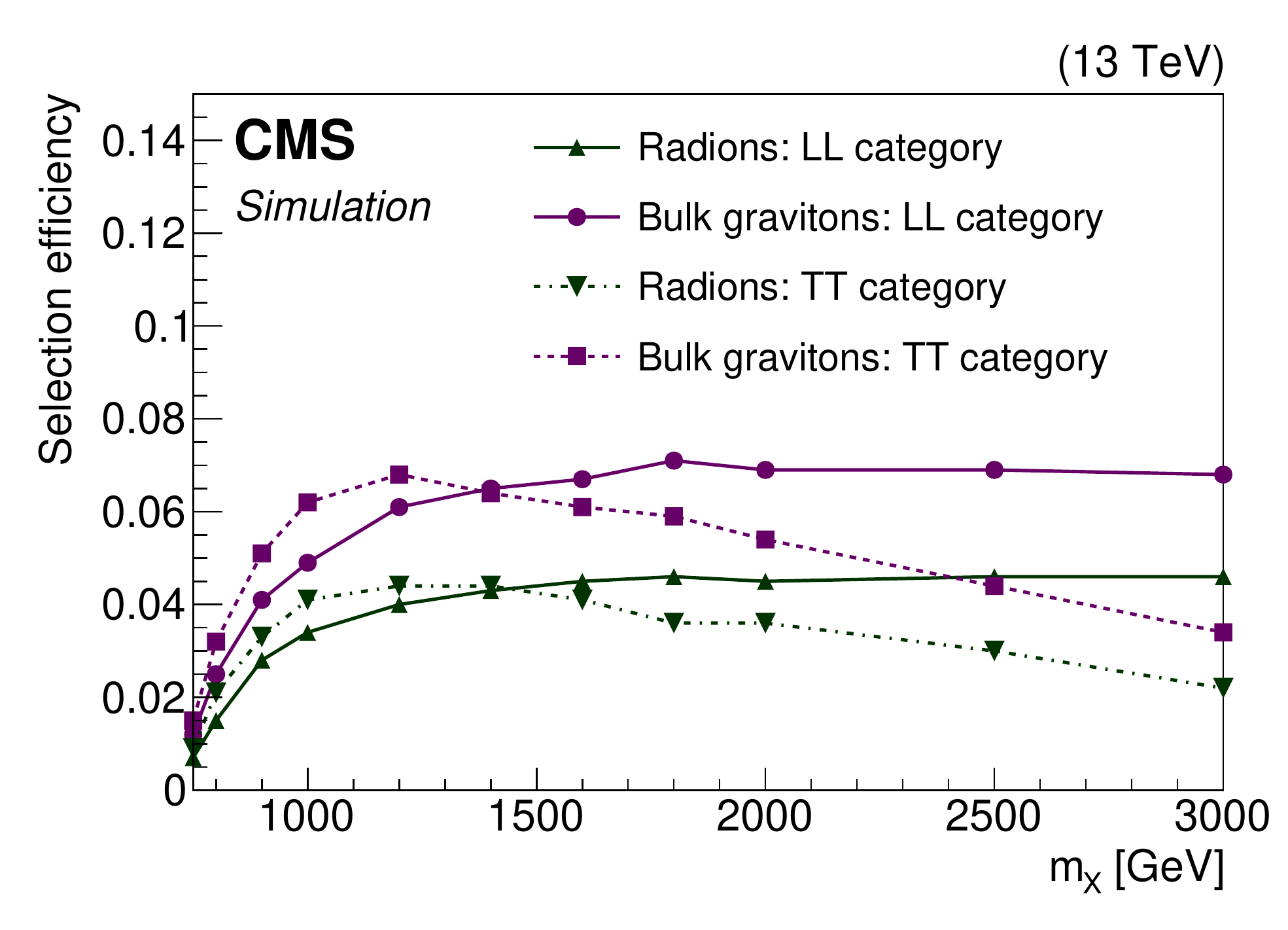}
\caption{The signal selection efficiencies for the bulk graviton and
  radion models for different mass hypotheses of the resonances, shown
  for the LL and the TT signal event categories. Owing to the large sample sizes of the simulated events, the statistical uncertainties are small.}
\label{fig:SignalEff}
\end{figure}

The backgrounds are estimated separately for each category, and the
combination of the likelihoods for the TT and LL categories gives
the optimal signal sensitivity over a wide range of resonance masses,
according to studies performed using
simulated signal and multijet samples. The TT category has
a good background rejection for $\mx$ up to 2000\GeV. At higher
resonance masses,
where the background is small, the LL category provides better
signal sensitivity. The full event selection efficiencies for bulk
gravitons and radions of different assumed masses are shown in
Fig.~\ref{fig:SignalEff}. The radion has a smaller efficiency than the
bulk graviton because its $\abs{\DeltaEta}$ distribution is
considerably wider than that of a bulk graviton of the same mass, as
shown in Fig.~\ref{fig:deta_mjjred_prebtag} (left).
\section{Signal and background modelling\label{sec:SigModelBkgEst}}

The method chosen for the background modelling depends on whether the resonance mass $\mx$ is below or above 1200\GeV, since at low masses the background does not fall smoothly as a function of \mjjs, because of the trigger requirements, while above 1200\GeV it does.
The background estimation relies on a set of control regions to predict the total background shape and normalization in the signal regions.
The entire range of the $\mjjs$ distribution above 750\GeV is used for the prediction.

\begin{figure}
\centering
\includegraphics[width=0.49\textwidth]{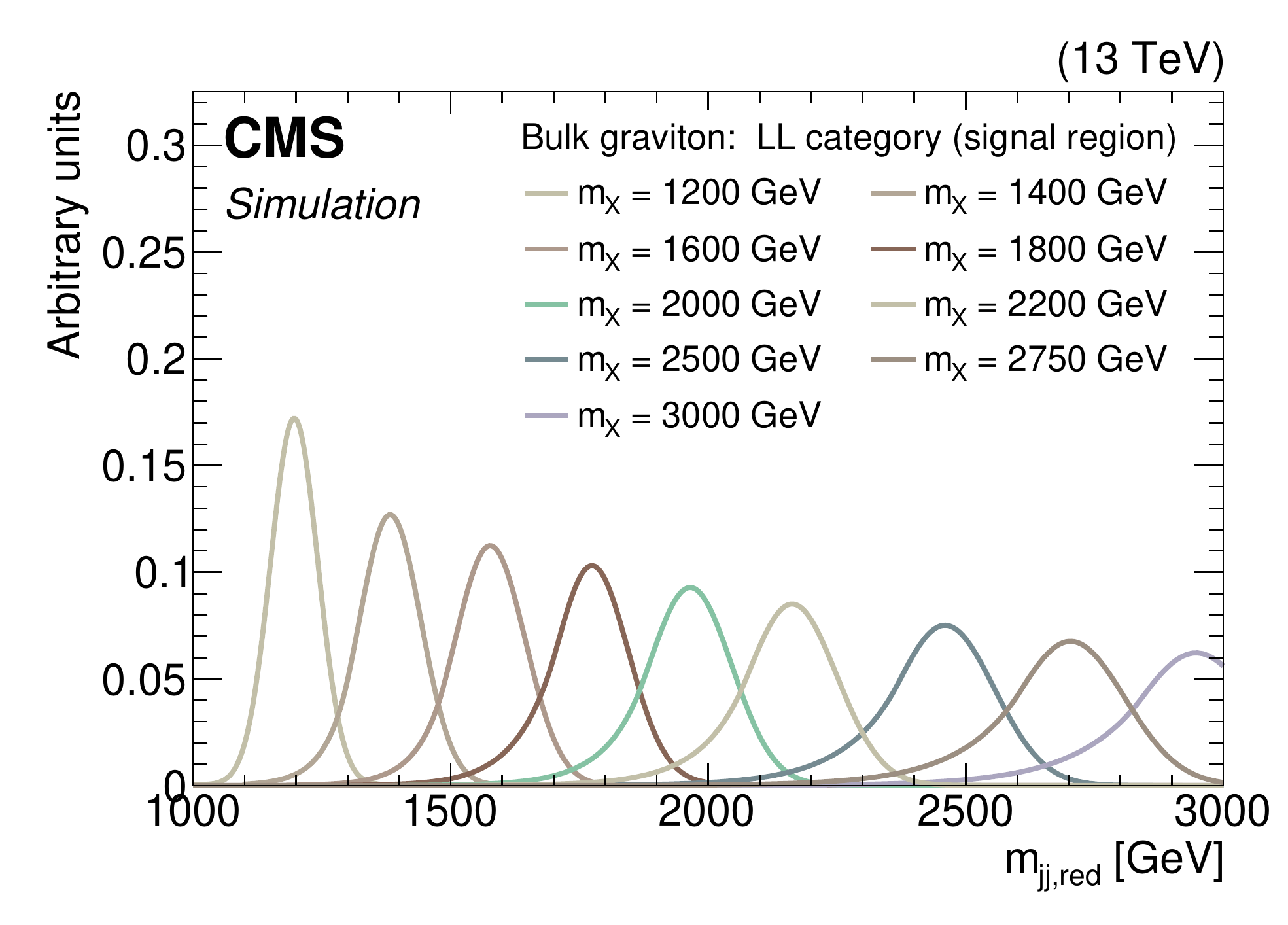}
\caption{The bulk graviton signal $\mjjs$ distribution
  for the LL category, modelled using the sum of Gaussian and Crystal Ball
  functions. This modelling is performed for signals in the range $1100 < \mjjs <
  3000\GeV$, where the background distribution falls smoothly. No events are observed with \mjjs greater than 3000\GeV.}
\label{fig:sigmodel}
\end{figure}

For signals with $\mx \ge 1200\GeV$, the underlying background distribution falls monotonically with \mjjs, thus allowing the background shape to be modelled by a smooth function, above which a localized signal is searched for. This smooth background modelling helps to reduce uncertainties in the background estimation from local statistical fluctuations in \mjjs, thereby improving the signal search sensitivity.
The parameters of the function and its total normalization are constrained by a simultaneous fit of the signal and background models to the data in the control and the signal regions.
For $\mx \ge 1200\GeV$, the $\mjjs$ distributions for the signal are modelled using the sum of
a Gaussian and a Crystal Ball function~\cite{Oreglia:1980cs},
as shown in Fig.~\ref{fig:sigmodel} for one signal category. The same modelling is used for the other signal categories, with different parameters for the Gaussian and the Crystal Ball functions.

\begin{table}[htbp]
  \topcaption{Definition of the signal, the antitag, and the sideband regions used for the background estimation. The regions are defined in terms of the soft-drop masses of the leading $\pt$ (j$_{1}$) and the subleading $\pt$ (j$_{2}$) AK8 jets, and their \Hbbt discriminator values.}
  \label{tab:EvtSel}
  \centering
  \setlength{\tabcolsep}{0.5em}
    \begin{tabular}{@{}lccc@{}}
      \hline\
      Event & Jet & Soft-drop&Double-\PQb\,tagger\\
       category& &  mass (\GeVns{}) &  discriminator\\\hline
      \multirow{2}{*}{Signal (LL)}   & j$_{1}$ & \multirow{4}{*}{105--135} & {$>$0.3, but} \\
                                            & j$_{2}$ & & {not both $>$0.8} \\
      \multirow{2}{*}{Signal (TT)}   & j$_{1}$ & & \multirow{2}{*}{$>$0.8} \\
                                            & j$_{2}$ & & \\[2.4ex]
                                            \hline
      \multirow{2}{*}{Antitag (LL)} & j$_{1}$ & \multirow{4}{*}{105--135} & $<$0.3 \\
                                            & j$_{2}$ & & 0.3--0.8 \\
      \multirow{2}{*}{Antitag (TT)} & j$_{1}$ & & $<$0.3 \\
                                            & j$_{2}$ & & $>$0.8 \\[2.4ex]
                                            \hline
      Sideband & j$_{1}$ & $<$105 or $>$135 &  {$>$0.3, but} \\
      (LL, passing) & j$_{2}$ & 105--135 & {not both $>$0.8} \\
      Sideband & j$_{1}$ & $<$105 or $>$135 & \multirow{2}{*}{$>$0.8} \\
      (TT, passing) & j$_{2}$ & 105--135 & \\[2.4ex]
                                            \hline
      Sideband & j$_{1}$ & $<$105 or $>$135 & $<$0.3 \\
      (LL, failing) & j$_{2}$ & 105--135 & 0.3--0.8 \\
      Sideband & j$_{1}$ & $<$105 or $>$135 & $<$0.3 \\
      (TT, failing) & j$_{2}$ & 105--135 & $>$0.8 \\
      \hline
    \end{tabular}
\end{table}

\begin{figure}[htb]
\centering
\includegraphics[width=0.49\textwidth]{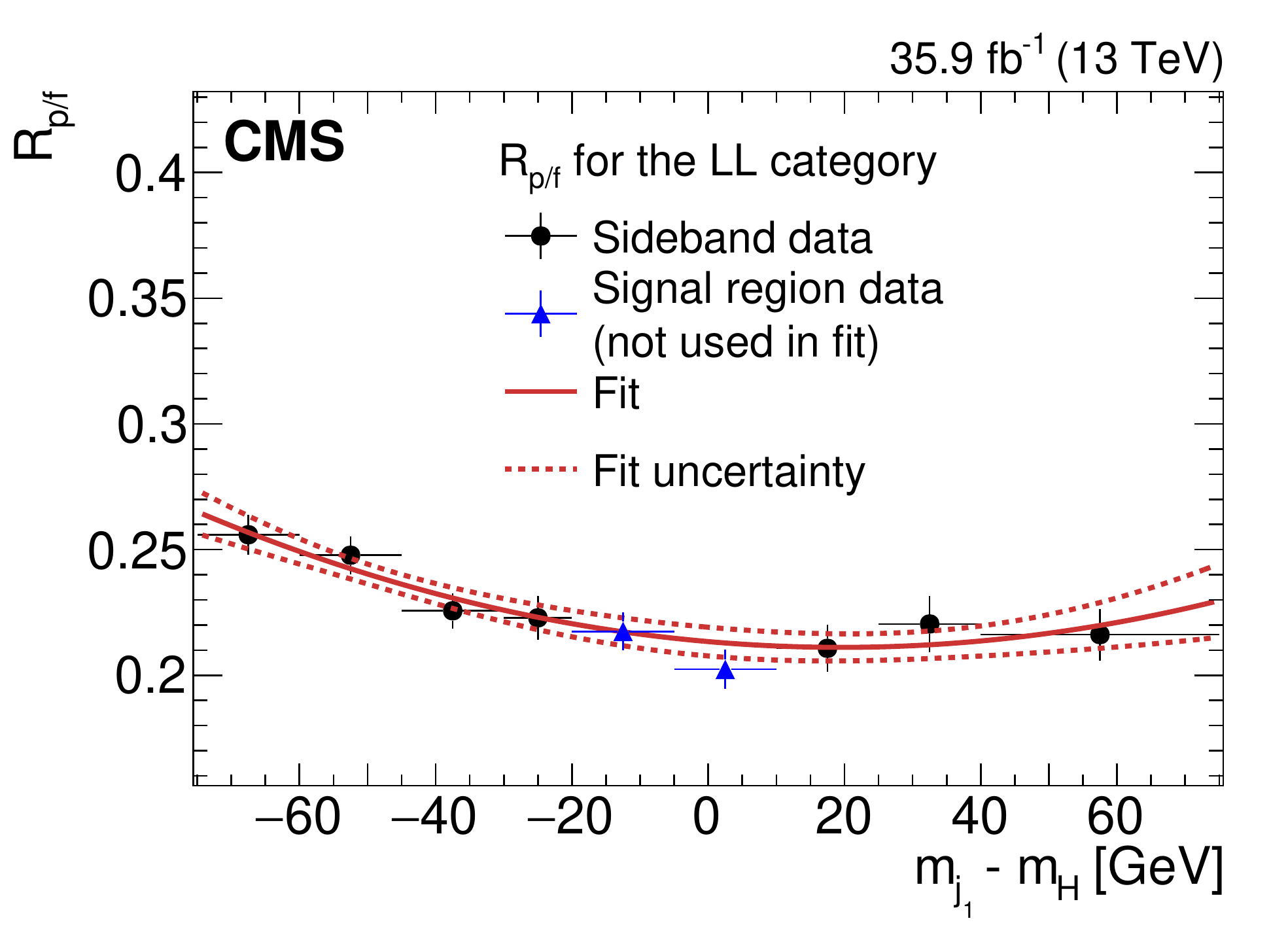}
\includegraphics[width=0.49\textwidth]{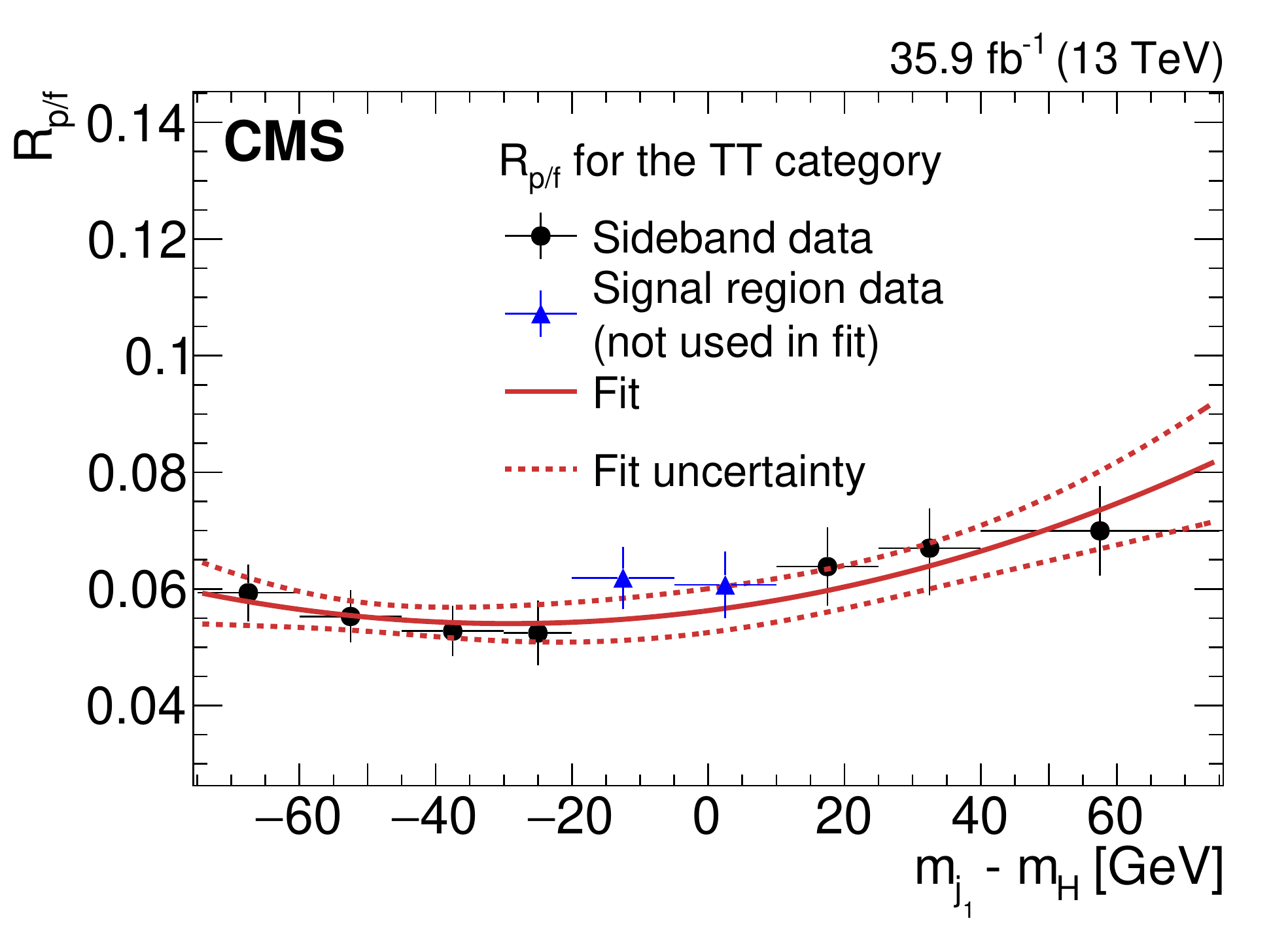}
\caption{The pass-fail ratio $R_\text{p/f}$ of the leading $\pt$ jet for the LL (\cmsLeft) and TT (\cmsRight) signal region categories as a function of the difference between the soft-drop mass of the leading jet and the Higgs boson mass, {\mjone - \mH}. The measured ratio in different bins of $\mjone - \mH$ is used in the fit (red solid line), except in the region around $\mjone - \mH = 0$, which corresponds to the signal region (blue triangular markers). The fitted function is interpolated to obtain $R_\text{p/f}$ in the signal region. The horizontal bars on the data points indicate the bin widths.}
\label{fig:Alphabet_Rpf_TT_LL}
\end{figure}

\begin{figure}[htbp]
\centering
\includegraphics[width=0.49\textwidth]{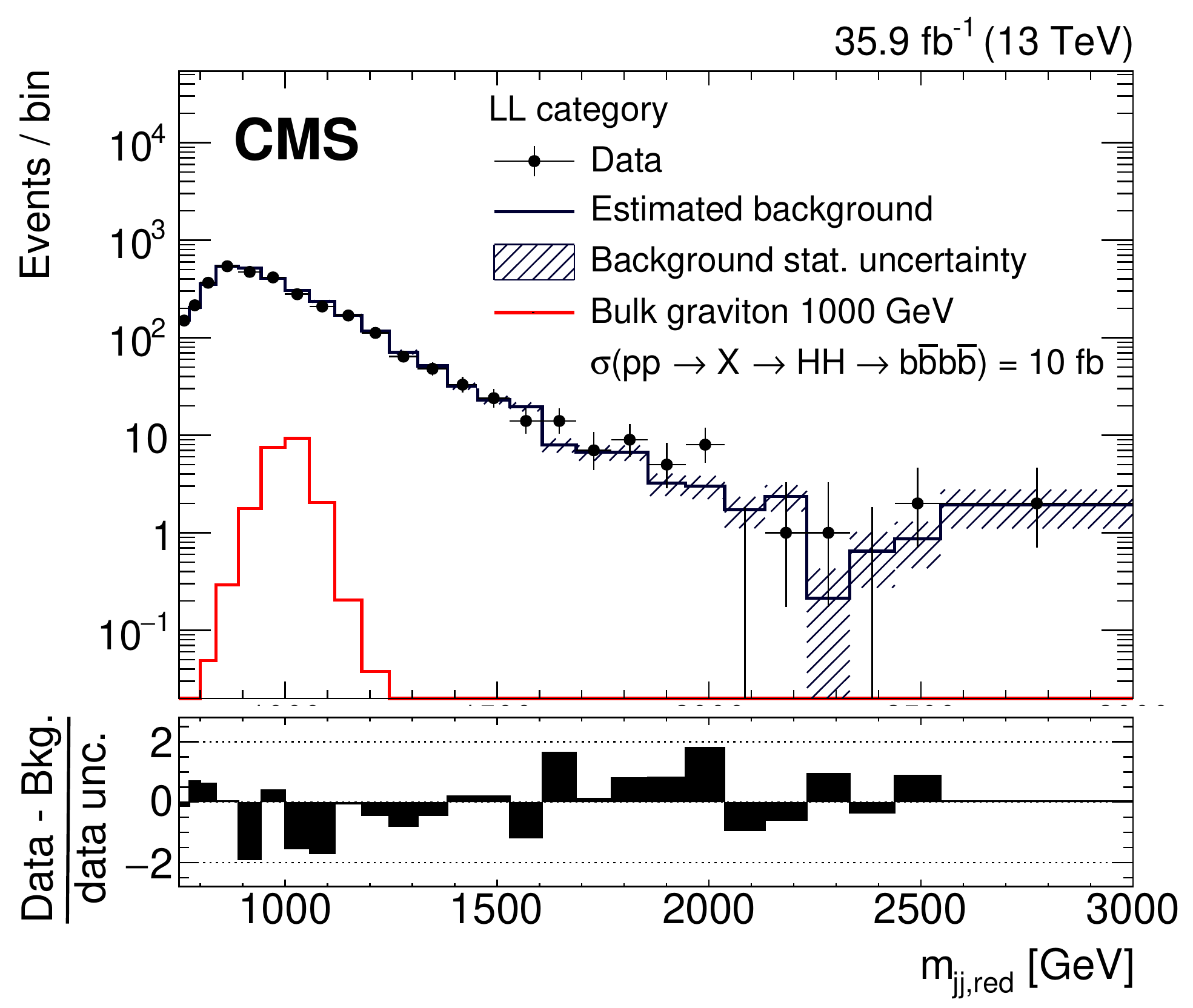}
\includegraphics[width=0.49\textwidth]{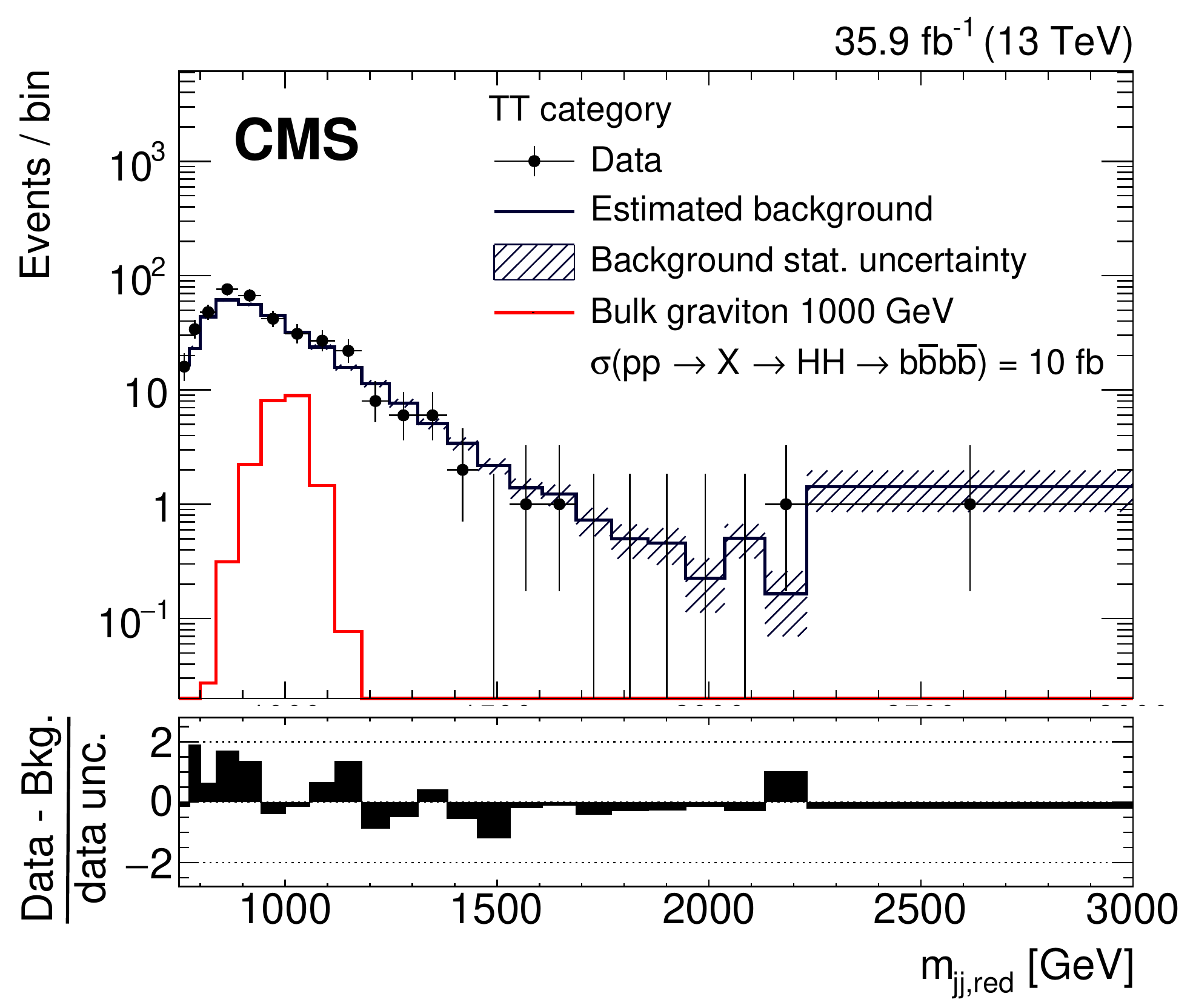}
\caption{The reduced mass distributions \mjjs for the LL (\cmsLeft) and TT (\cmsRight) signal region categories. The points with bars show the data, the histogram with shaded band shows the estimated background and associated uncertainty. The \mjjs spectrum for the background is obtained by weighting the \mjjs spectrum in the antitag region by the ratio $R_\text{p/f}$  of Fig.~\ref{fig:Alphabet_Rpf_TT_LL}. The signal predictions for a bulk graviton of mass 1000\GeV, are overlaid for comparison, assuming a cross section $\sigma(\Pp\Pp \to X \to \PH\PH \to \bbbar\bbbar)$ of 10\unit{fb}. The last bins of the distributions contain all events with $\mjjs > 3000\GeV$. The differences between the data and the predicted background, divided by the data statistical uncertainty (data unc.) as given by the Garwood interval~\cite{Garwood}, are shown in the lower panels.}
\label{fig:Alphabet_Bkg_TT_LL}
\end{figure}

\begin{figure*}
\centering
\includegraphics[width=0.49\textwidth]{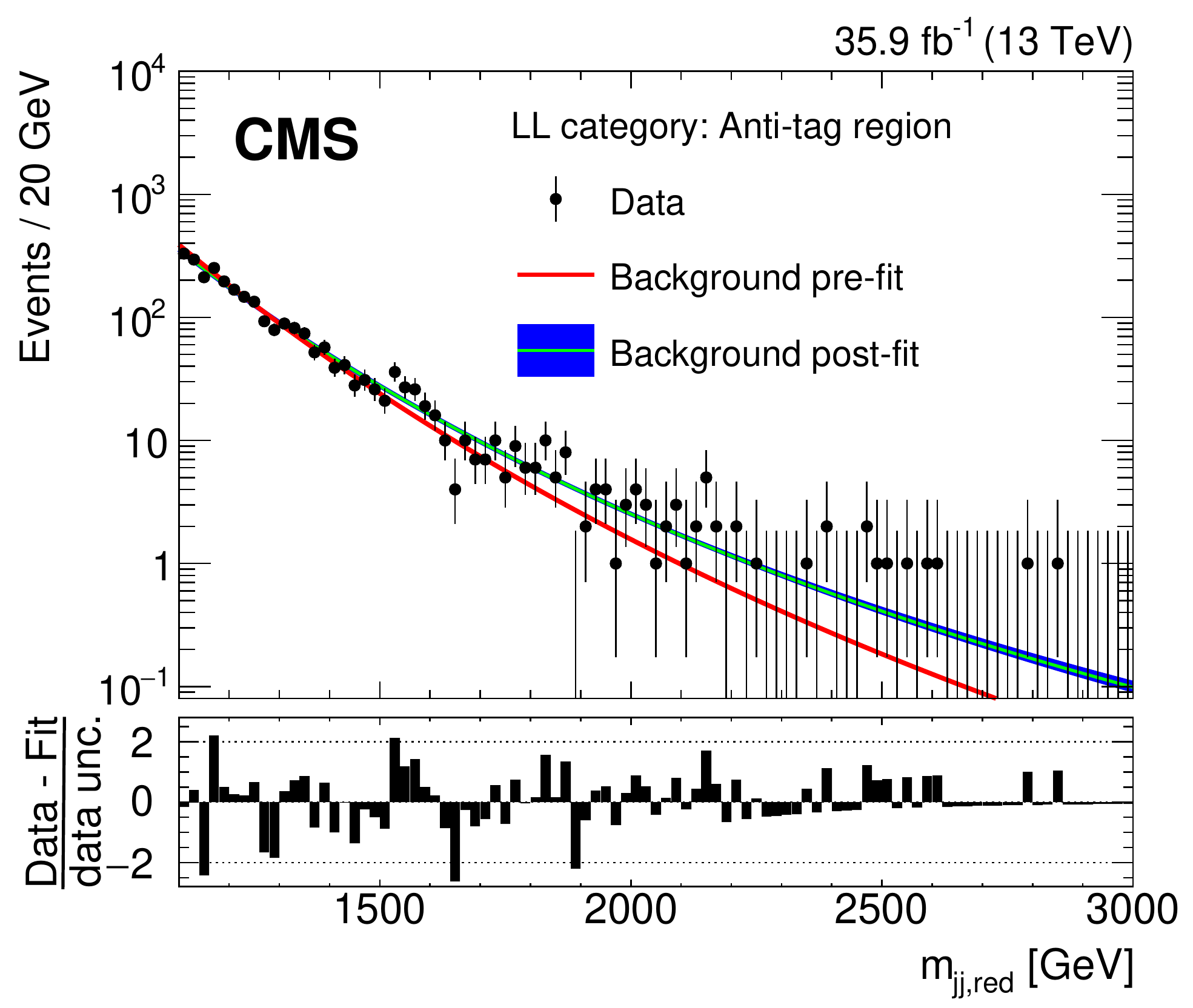}
\includegraphics[width=0.49\textwidth]{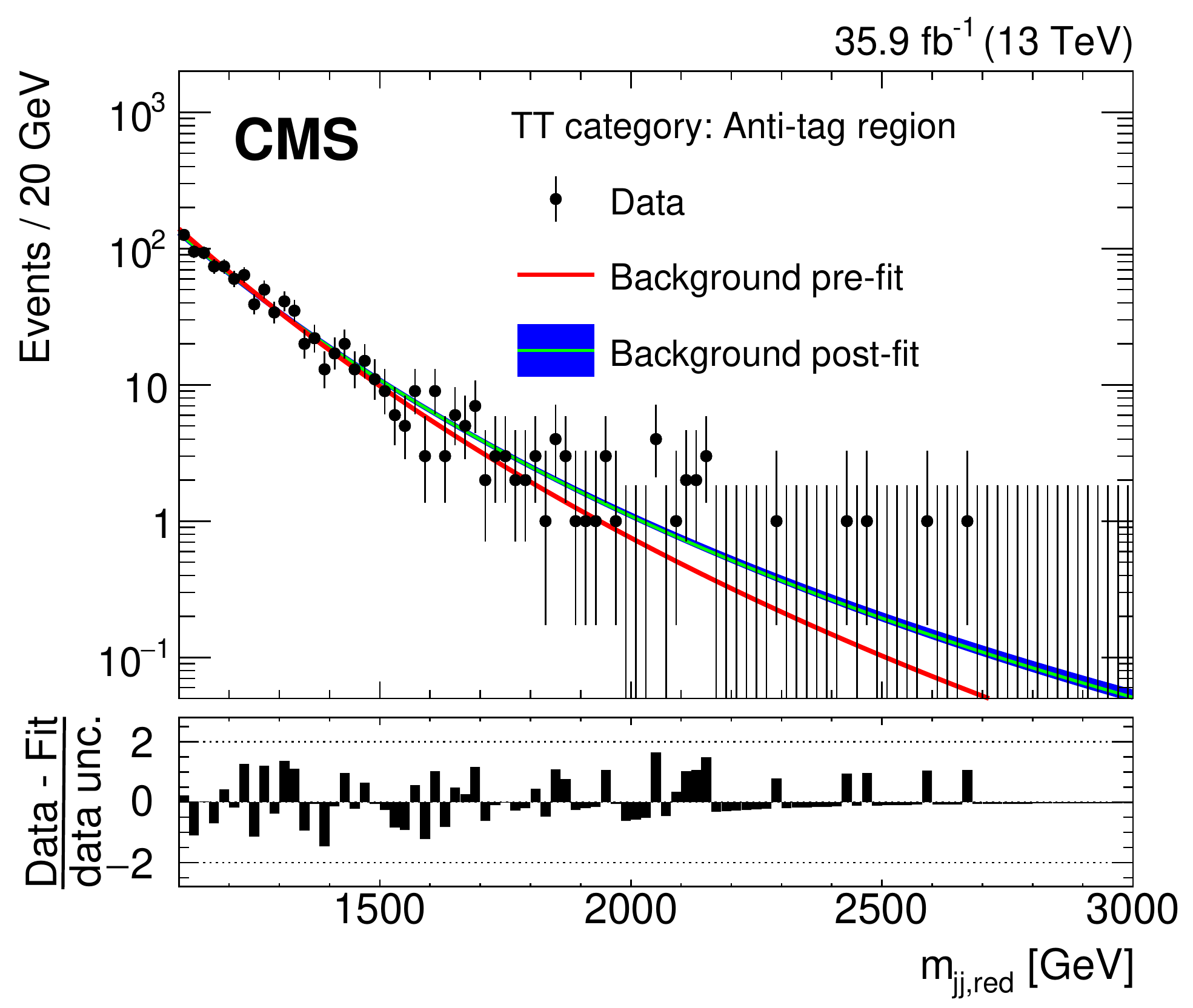}
  \caption{The reduced mass \mjjs distributions in the antitag region for the LL
    (left) and TT (right) categories. The black markers are the data
    while the curves show the pre-fit and post-fit background
    shapes.
    The differences between the data and the pre-fit
  background distribution, divided by the statistical uncertainty in the data
  (data unc.) as given by the Garwood interval~\cite{Garwood}, are shown in the lower panels.}
\label{fig:AABH_AT}
\end{figure*}

\begin{figure*}
\centering
\includegraphics[width=0.49\textwidth]{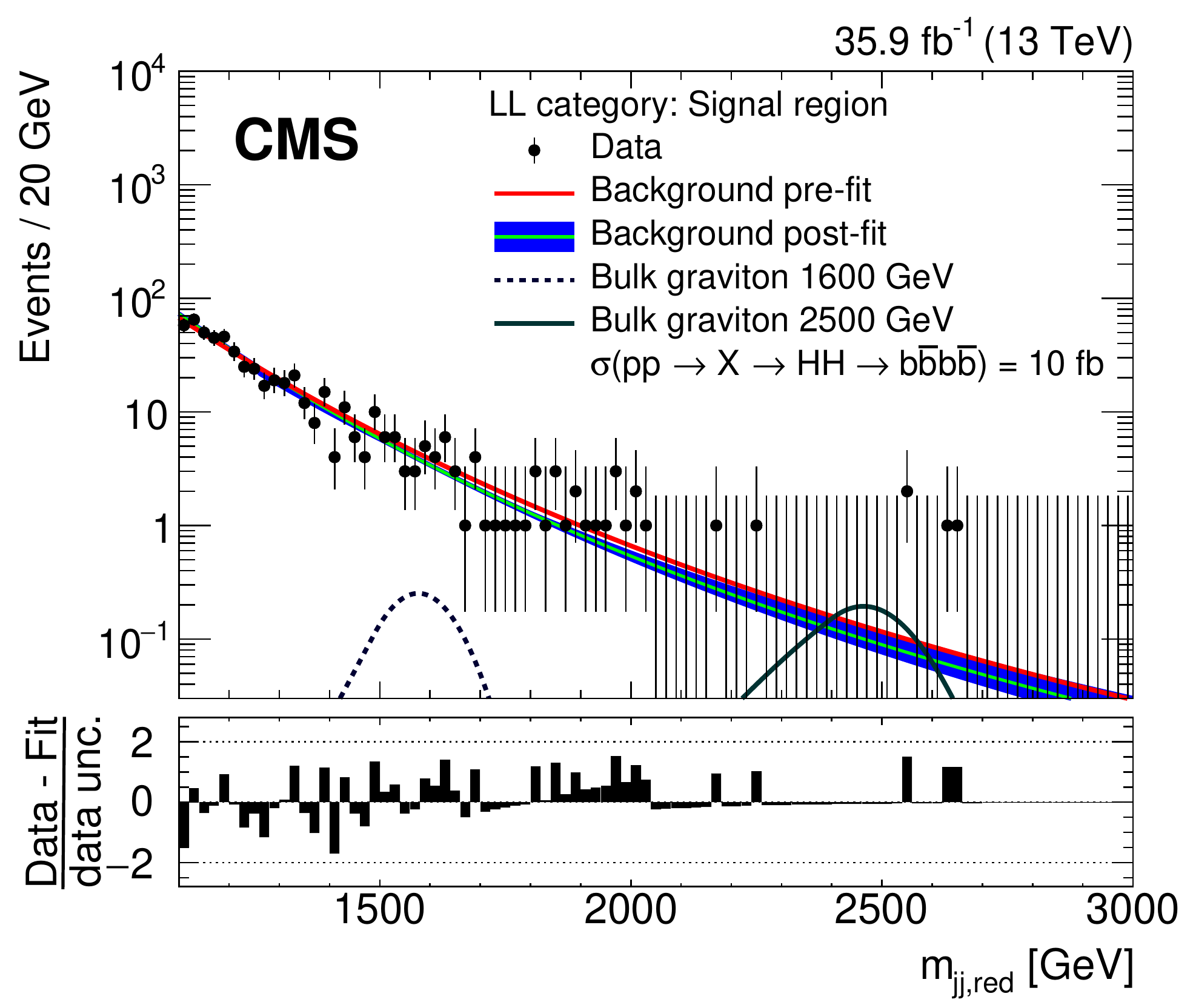}
\includegraphics[width=0.49\textwidth]{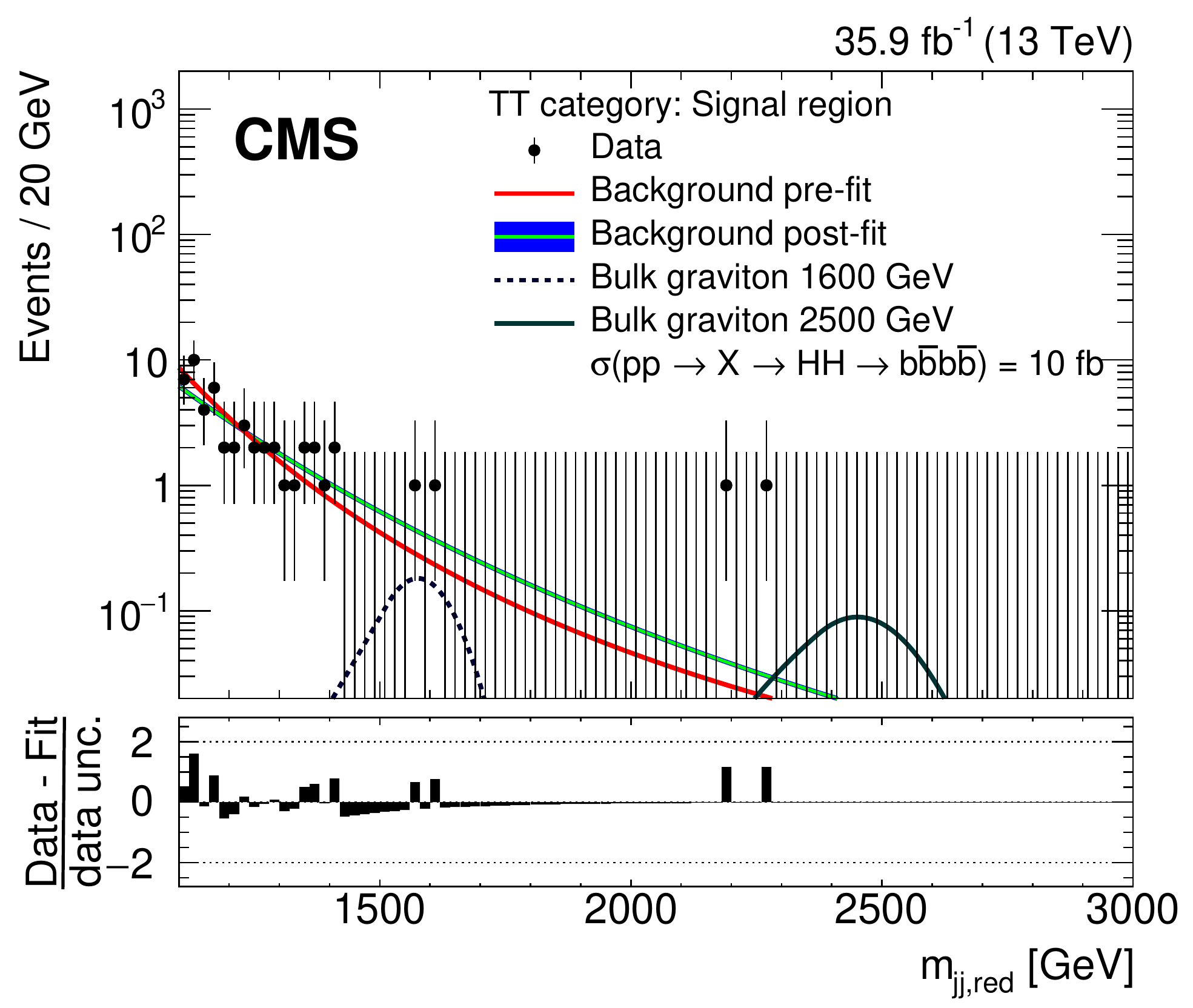}
\caption{The reduced mass \mjjs distributions in the signal region for the LL
  (left) and the TT (right) categories. The black markers are the data
  while the curves show the pre-fit and post-fit background
  shapes. The contribution of bulk gravitons of masses 1600 and
  2500\GeV in the signal region are shown assuming a cross section $\sigma(\Pp\Pp \to X \to \PH\PH \to \bbbar\bbbar)$ of 10\unit{fb}.
  The differences between the data and the pre-fit
  background distribution, divided by the statistical uncertainty in the data
  (data unc.) as given by the Garwood interval~\cite{Garwood}, are shown in the lower panels.
  }
\label{fig:AABH_SR}
\end{figure*}

The signal and control regions are defined by two variables related to the leading $\pt$ jet j$_{1}$: (i) its soft-drop mass $\mjone$ and (ii) the value of the discriminator of the \Hbbt. The background is estimated in bins
of the \mjjs distribution. Considering these two variables, several regions are defined.

The \textit{pre-tag} region includes events fulfilling the selection requirements in Sections~\ref{sec:CMSDetector}--\ref{sec:EvtSel} apart from those on \mjone and on the j$_{1}$ \Hbbt discriminator.
The \textit{signal} region is the subset of pre-tag events where \mjone is inside the $\PH$ jet mass window of 105--135\GeV, and with the j$_{1}$ \Hbbt discriminator greater than 0.3 or 0.8, for the LL and TT regions, respectively.
The \textit{antitag} regions require the j$_{1}$ \Hbbt discriminator to be less than 0.3, with the requirement on  j$_{2}$ being the same as that for the corresponding LL or TT signal regions.
The \mjone \textit{sideband} region consists of events in the pre-tag region, where
\mjone lies outside the $\PH$ jet mass window. Based on whether  j$_{1}$ passes or fails the \Hbbt discriminator threshold, the sideband region is divided into either ``passing'' or ``failing'', respectively.
The antitag regions are dominated by the multijet background, and have identical kinematic distributions to the multijet background events in the signal region, according to studies using simulations.
The definitions of the signal, the antitag, and the sideband regions are given in Table~\ref{tab:EvtSel}.

In the absence of a correlation between \mjone and the \Hbbt
discriminator values, one could measure in the \mjone sideband the
ratio of the number of events passing and failing the \Hbbt selection,
$R_\text{p/f} \equiv N_\text{pass}/N_\text{fail}$, \ie the ``pass-fail ratio".
The yield in the antitag region (in each $\mjjs$ bin) could then be scaled by $R_\text{p/f}$ to obtain an estimate of the background normalization in the signal region. However, there is a small correlation between the \Hbbt discriminator and \mjone, which is taken into account by measuring the pass-fail ratio $R_\text{p/f}$ as a function of \mjone.
The signal fraction was found to be less than $10^{-3}$ in the sideband regions used to evaluate $R_\text{p/f}$, assuming a signal cross section $\sigma(\Pp\Pp \to X \to \PH\PH \to \bbbar\bbbar)$ of $10\unit{fb}$.

The $R_\text{p/f}$ for the LL signal region is measured using ratio of the number of events in the ``LL, passing'' and ``LL, failing'' sideband regions, as defined in Table~\ref{tab:EvtSel}. Likewise, the $R_\text{p/f}$ for the TT signal region uses the ratio of the number of events in the ``TT, passing'' to the ``TT, failing'' sideband regions.
The variation of $R_\text{p/f}$ as a function of \mjone in each \mjone sideband is fitted with a quadratic function.
The fit to the pass-fail ratio is interpolated to the region where \mjone lies within the $\PH$ jet mass window of 105--135\GeV. An alternative fit using a third order polynomial was found to give the same interpolated value of $R_\text{p/f}$ in the Higgs jet mass window. Every event in the antitag region is scaled by the pass-fail ratio evaluated for the $\mjone$ of that event, to obtain the background prediction in the signal region.

Figure~\ref{fig:Alphabet_Rpf_TT_LL} shows the quadratic fit in the
\mjone sidebands of the pass-fail ratio $R_\text{p/f}$ as a function of
\mjone, as obtained in the data. The background prediction using this
method, along with the number of observed events in the signal region
is shown in Fig.~\ref{fig:Alphabet_Bkg_TT_LL}.

For resonance masses of 1200\GeV and above, the background estimation is
improved by simultaneously fitting a parametric model for the
background and signal to the data in the signal and the antitag
regions for $\mjjs \ge 1100\GeV$. In the fit, the ratio $R_\text{p/f}$ obtained from the sidebands is used to constrain the relative number of background events in the two regions.
To account for possible $R_\text{p/f}$ dependence on \mjjs at high \mjjs values,
the $R_\text{p/f}$ obtained from the fits shown in Fig.~\ref{fig:Alphabet_Rpf_TT_LL} is also parametrized as a linear function of \mjjs.
The signal normalization is unconstrained in
the fit, while the uncertainties in the parameters of the functions
used to model the background and $R_\text{p/f}$ are treated as nuisance
parameters.
For the background modelling, a choice among an exponential function $N \re^{-a\, \mjjs}$, a ``levelled exponential'' function $N \re^{-a\,\mjjs/(1 + a\, b\, \mjjs)}$, and a ``quadratic levelled exponential function'' $N \re^{[{-a\, \mjjs}/(1 + a\, b\, \mjjs)] - [{-c\,\mjjs^{2}}/(1 + b\, c\, \mjjs^{2})]}$ was made, using a Fisher F-test~\cite{Fisher}. At a confidence level of 95\%, the levelled exponential function was found to be optimal.
Since the background shapes in the signal regions, as predicted using the antitag regions, were found to be similar (Fig.~\ref{fig:Alphabet_Bkg_TT_LL}), the parametric background modelling was tested using the antitag region in the data before applying it to the signal region.

The simultaneous fits to the antitag and the signal regions are shown in Figs.~\ref{fig:AABH_AT} and \ref{fig:AABH_SR}, respectively, using the background model only. These are labelled as ``post-fit'' curves with the signal region background yields constrained to be $R_\text{p/f}$ times the background yields from the antitag regions. The ``pre-fit'' curves, obtained by fitting the antitag and the signal regions separately to the background-only model, with the background event yields unconstrained, are also shown for comparison. In the post-fit results, the $R_\text{p/f}$ dependence on \mjjs was found to be negligible.

Among the four fitted regions, corresponding to the antitag and the signal regions in the
LL and TT categories, the events with the highest value of \mjjs
occur in the antitag region of the LL category, at around
$\mjjs = 2850\GeV$.
As the parametric background model is only reliable within the range
of observed events, the likelihood is only evaluated up to $\mjjs = 3000\GeV$.
This results in a truncation of the signal distribution for resonances having $\mx$ of 2800\GeV and above, with signal efficiency losses increasing to 30\% for $\mx = 3000\GeV$, as shown in Fig.~\ref{fig:sigmodel}.

Closure tests of the background estimation methods were performed using simulated multijet samples with signals of various cross sections. The tests indicated a good consistency between the expected and the assumed signal strengths.

\section{Systematic uncertainties\label{sec:Systematics}}

The following sources of systematic uncertainty affect the expected
signal yields. None of these lead to a
significant change in the signal shape.

Trigger response modelling uncertainties are
particularly important for $\mjjs < 1200\GeV$, where the trigger
efficiency drops below 99\%. A scale factor is applied to correct for
the difference in efficiency observed between the data and simulation.
The control trigger used to measure this scale factor requires a single AK4 jet with $\pt > 260\GeV$, and it too is subject to some inefficiency when \mjjs is close to 750\GeV, because of a difference between the jet energy scale used in the trigger and that used in the offline reconstruction. This inefficiency is measured using simulations, and has an associated total uncertainty of between 1\% and 15\%.

\begin{table}[htb]
  \topcaption{Summary of systematic uncertainties in the signal and background yields.}
  \label{tab:Syst}
  \centering
    \begin{tabular}{l{c}@{\hspace*{5pt}}c}
      \hline
      Source && \multicolumn{1}{r}{Uncertainty (\%)} \\ \hline \\[-2ex]
      \multicolumn{3}{c}{Signal yield} \\[1.8ex]

      Trigger efficiency                    && 1--15        \\
      $\PH$ jet energy scale and resolution && 1            \\
      $\PH$ jet mass scale and resolution   && 2            \\
      $\PH$ jet $\nsub$ selection           && $+$30 /$-$26    \\
      $\PH$-tagging correction factor       && 7--20        \\
      Double-\PQb\,tagger discriminator        & & 2--5         \\
      Pileup modelling                      && 2            \\
      PDF and scales                        && 0.1-2        \\
      Luminosity                            && 2.5          \\
      \multicolumn{3}{c}{Background yield} \\[1.8ex]
      \multicolumn{3}{l}{$R_\text{p/f}$ fit\hfill                    2.6 (LL category) 6.8 (TT category)} \\
      \hline
    \end{tabular}
\end{table}

The jet energy scale and resolution uncertainty is about 1\%~\cite{Khachatryan:2016kdb,CMS-DP-2016-020}.
The jet mass scale and resolution, and $\tau_{21}$ selection efficiency data-to-simulation scale factor are measured using a sample of merged $\PW$ jets in semileptonic $\ttbar$ events. The corresponding uncertainties are extrapolated to a higher $\pt$ range than that associated with $\ttbar$ events, using simulations.
A correction factor is applied to account for the difference in the jet shower profile of $\PW \to {\qqbar'}$ and $\Hbb$ decays, by comparing the ratio of the efficiency of $\PH$ and $\PW$ jets using the \PYTHIA~8 and \HERWIGpp shower generators.
The jet mass scale and resolution has a 2\% effect on the signal yields because of a change in the mean of the $\PH$ jet mass distribution. The $\nsub$ selection efficiency uncertainty amounts to a ${+}30{/}{-}26\%$ change in the signal yields. The uncertainty in the $\PH$ tagging correction factor is in the range 7--20\% depending on the resonance mass $\mx$. The \Hbbt efficiency scale factor uncertainty is about 2--5\%, depending on the \Hbbt requirement threshold and jet \pt, and is propagated to the total uncertainty in the signal yield.

The impact of the PDFs and the theoretical
scale uncertainties are estimated to be 0.1--2\%, using the {PDF4LHC}
procedure~\cite{Butterworth:2015oua}, and
affect the product of the signal acceptance and the efficiency.
The PDF and scale uncertainties have negligible impact on the signal $\mjjred$ distributions.
Additional systematic uncertainties associated with the pileup modelling (2\%) and the integrated luminosity determination (2.5\%)~\cite{CMS-PAS-LUM-17-001}, are applied to the signal yield.

The main source of uncertainty for the multijet background in the
region $\mjjs < 1200\GeV$ is due to the statistical uncertainty in the
fit to the $R_\text{p/f}$ ratio performed in the $\PH$ jet mass
sidebands. This uncertainty, amounting to 2.6\% for the LL, and 6.8\% for the TT signal categories, is fully correlated between all bins of a
particular estimate. Furthermore, the statistical
uncertainty in the antitag region is propagated to the signal
region when the estimate is made. This is uncorrelated from bin to bin, and the
Barlow--Beeston Lite~\cite{BarlowBeeston,BBLite} method is used to treat the
bin-by-bin statistical uncertainty in the data. These uncertainties affect both
the shape of the background in the \mjjs distribution and the total
background yield.

For $\mjjs \ge 1200\GeV$, the overall background uncertainty is obtained from the uncertainty in the four simultaneous fits performed for the antitag and the signal regions in the LL and the TT categories. The dependence of $R_\text{p/f}$ on \mjjs is accounted for, although this was found to be negligible.

A complete list of systematic uncertainties is given in Table~\ref{tab:Syst}.
\section{Results\label{sec:Results}}

\begin{figure}[htb]
  \centering
  \includegraphics[width=0.49\textwidth]{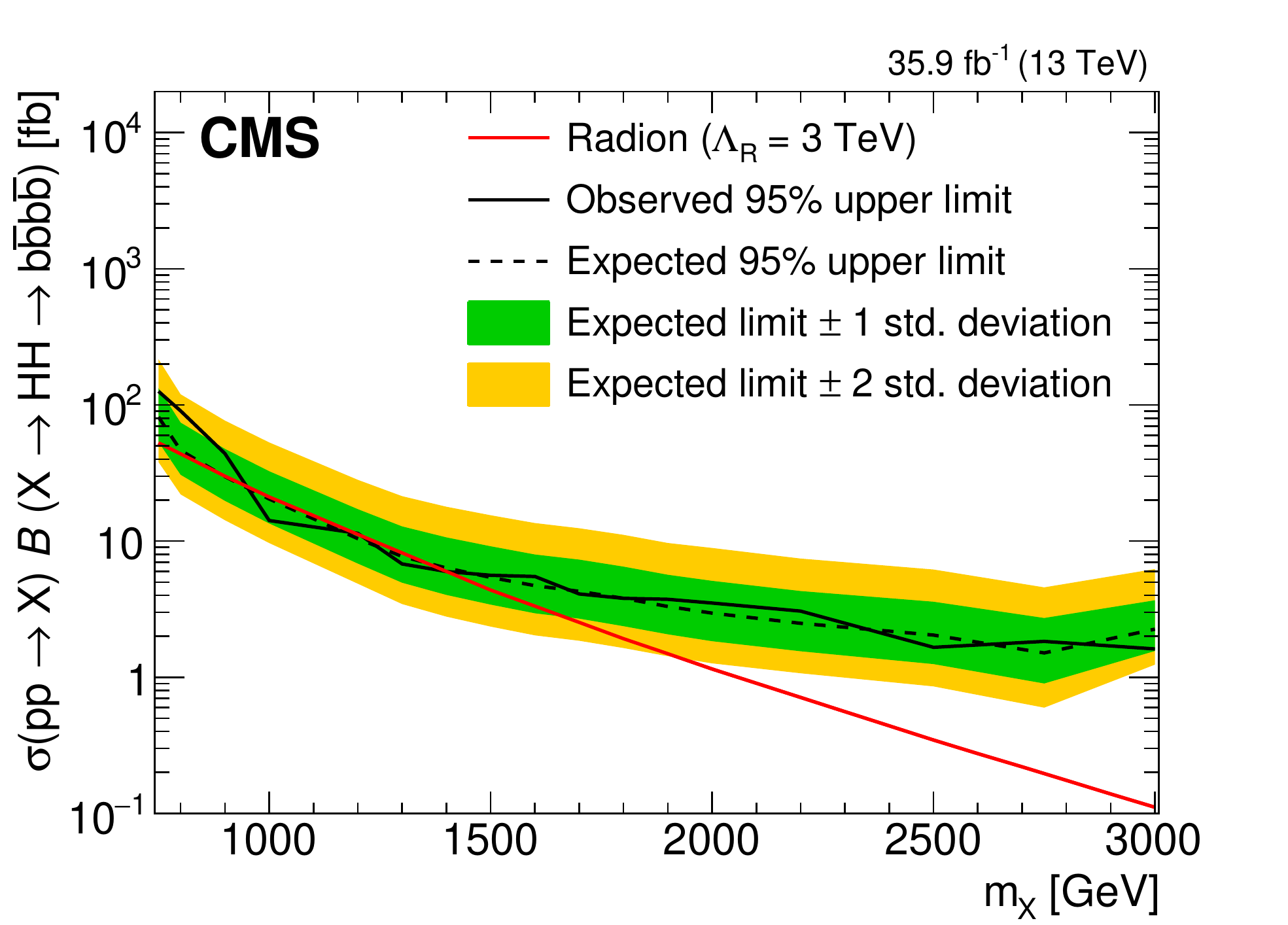}
  \includegraphics[width=0.49\textwidth]{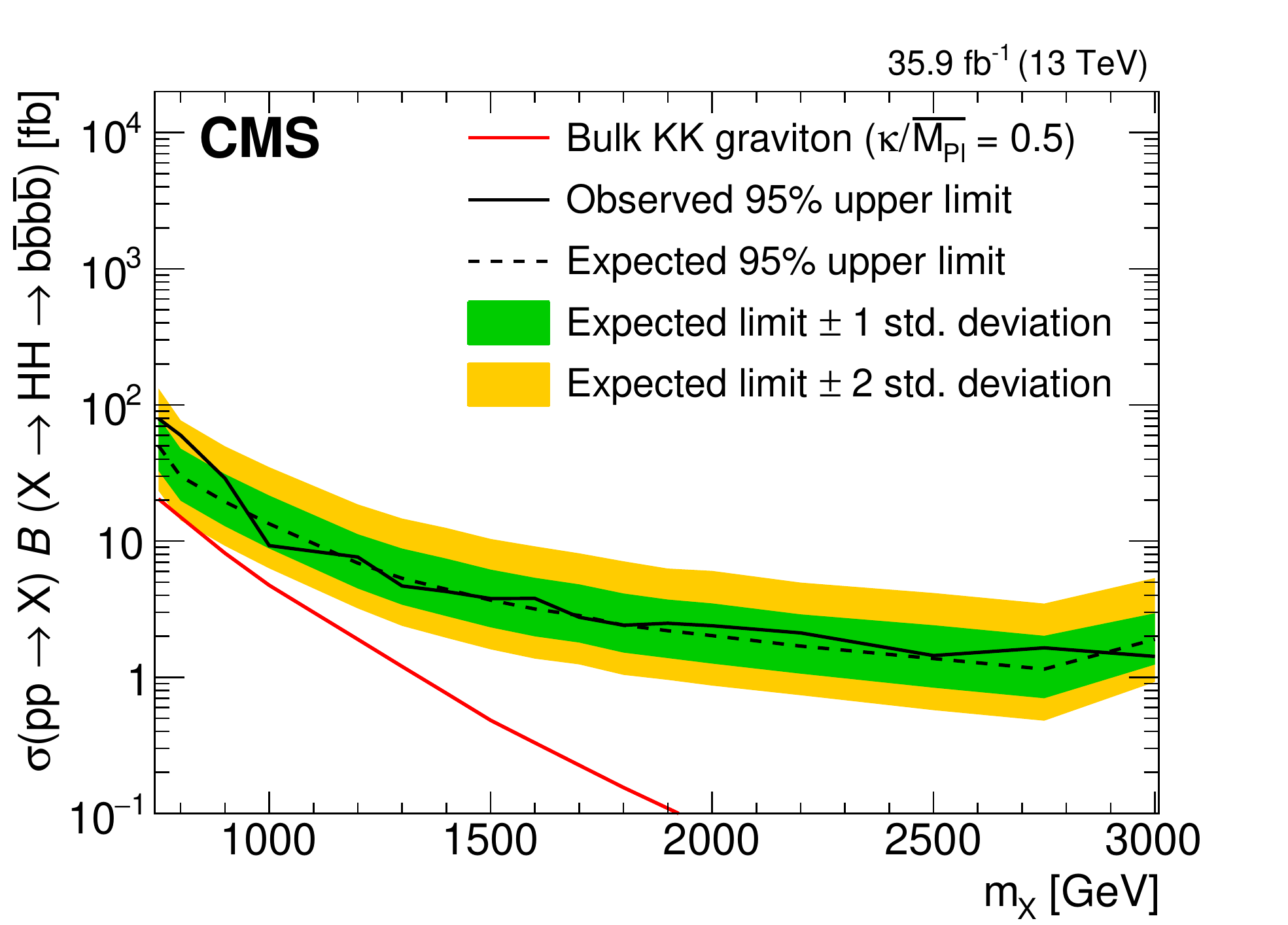}
  \caption{The limits for the spin-0 radion (\cmsLeft) and the
    spin-2 bulk graviton (\cmsRight) models. The result for $\mx < 1200\GeV$ uses the background predicted using the control regions, while for $\mx \ge 1200\GeV$ the background is derived from a combined signal and background fit to the data in the control and the signal regions. The
    predicted theoretical cross sections for a narrow radion or a bulk
    graviton are also
    shown.}
  \label{fig:lim_rad_grav_comb}
\end{figure}

As shown in Figs.~\ref{fig:Alphabet_Bkg_TT_LL}
and \ref{fig:AABH_SR}, for the signal regions, the observed \mjjs distribution is consistent with the estimated background.
The results are interpreted in terms of upper limits on the
product of the production cross sections and the branching fractions
$\sigma(\Pp\Pp \to \PX) \mathcal{B}(\PX \to \PH\PH \to \bbbar \bbbar)$
for radion and bulk graviton of various mass hypotheses.
The asymptotic approximation of the modified frequentist approach for confidence levels, taking the profile likelihood as a test
statistic~\cite{CLS1,CLS2,AsympCLs}, is used.
The limits are shown in Fig.~\ref{fig:lim_rad_grav_comb} for  a narrow width radion or a bulk
graviton. These are compared with the theoretical values of the product of the cross sections and branching fractions for the benchmarks $\kappa/\overline{\Mpl} = 0.5$ and $\LambdaR = 3\TeV$, where the narrow width approximation for a signal is valid, and where the corresponding $\PH\PH$ decay branching fractions in the mass range of interest are 10 and 23\%, for the graviton and the radion, respectively~\cite{Oliveira:2014kla}. The expected limits on the bulk graviton are more stringent than those on the radion because of the higher efficiency of the $\abs{\DeltaEta}$ separation requirement for the former signal.

The upper limits on the production of the cross sections and branching fraction lies in the range 126--1.4\unit{fb} for a narrow resonance $\PX$ of mass  $750 < \mx <  3000\GeV$. Assuming $\LambdaR = 3\TeV$, a bulk radion with a mass between 970 and 1400\GeV is excluded at 95\% confidence level, except in a small region close to 1200\GeV, where the observed limit is 11.4\unit{pb}, the theoretical prediction being 11.2\unit{pb}.

\section{Summary\label{sec:Summary}}

A search for a narrow massive resonance decaying to two standard model
Higgs bosons is performed using the LHC proton-proton collision data
collected at a centre-of-mass energy of 13\TeV by the CMS detector,
and corresponding to an integrated luminosity of \intLumi.
The final state consists of events with both Higgs bosons decaying to $\PQb$ quark-antiquark pairs, which were identified using jet substructure and $\PQb$-tagging techniques applied to large-area jets.
The data are found to be consistent with the standard model expectations, dominated by multijet events.
Upper limits are set on the products of the resonant production cross sections of a Kaluza--Klein bulk graviton and a Randall--Sundrum radion, and their branching fraction to $\PH\PH \to \bbbar\bbbar$.
The limits range from 126 to 1.4\unit{fb} at 95\% confidence level for bulk gravitons and radions in the mass range 750--3000\GeV.
For the mass scale $\LambdaR = 3\TeV$, a radion of mass between 970 and 1400\GeV (except in a small region close to 1200\GeV) is excluded.
These limits on the bulk graviton and the radion decaying to a pair of standard model Higgs bosons are the most stringent to date, over the mass range explored.

\begin{acknowledgments}
We congratulate our colleagues in the CERN accelerator departments for the excellent performance of the LHC and thank the technical and administrative staffs at CERN and at other CMS institutes for their contributions to the success of the CMS effort. In addition, we gratefully acknowledge the computing centres and personnel of the Worldwide LHC Computing Grid for delivering so effectively the computing infrastructure essential to our analyses. Finally, we acknowledge the enduring support for the construction and operation of the LHC and the CMS detector provided by the following funding agencies: BMWFW and FWF (Austria); FNRS and FWO (Belgium); CNPq, CAPES, FAPERJ, and FAPESP (Brazil); MES (Bulgaria); CERN; CAS, MoST, and NSFC (China); COLCIENCIAS (Colombia); MSES and CSF (Croatia); RPF (Cyprus); SENESCYT (Ecuador); MoER, ERC IUT, and ERDF (Estonia); Academy of Finland, MEC, and HIP (Finland); CEA and CNRS/IN2P3 (France); BMBF, DFG, and HGF (Germany); GSRT (Greece); OTKA and NIH (Hungary); DAE and DST (India); IPM (Iran); SFI (Ireland); INFN (Italy); MSIP and NRF (Republic of Korea); LAS (Lithuania); MOE and UM (Malaysia); BUAP, CINVESTAV, CONACYT, LNS, SEP, and UASLP-FAI (Mexico); MBIE (New Zealand); PAEC (Pakistan); MSHE and NSC (Poland); FCT (Portugal); JINR (Dubna); MON, RosAtom, RAS, RFBR and RAEP (Russia); MESTD (Serbia); SEIDI, CPAN, PCTI and FEDER (Spain); Swiss Funding Agencies (Switzerland); MST (Taipei); ThEPCenter, IPST, STAR, and NSTDA (Thailand); TUBITAK and TAEK (Turkey); NASU and SFFR (Ukraine); STFC (United Kingdom); DOE and NSF (USA).

\hyphenation{Rachada-pisek} Individuals have received support from the Marie-Curie programme and the European Research Council and Horizon 2020 Grant, contract No. 675440 (European Union); the Leventis Foundation; the A. P. Sloan Foundation; the Alexander von Humboldt Foundation; the Belgian Federal Science Policy Office; the Fonds pour la Formation \`a la Recherche dans l'Industrie et dans l'Agriculture (FRIA-Belgium); the Agentschap voor Innovatie door Wetenschap en Technologie (IWT-Belgium); the Ministry of Education, Youth and Sports (MEYS) of the Czech Republic; the Council of Science and Industrial Research, India; the HOMING PLUS programme of the Foundation for Polish Science, cofinanced from European Union, Regional Development Fund, the Mobility Plus programme of the Ministry of Science and Higher Education, the National Science Center (Poland), contracts Harmonia 2014/14/M/ST2/00428, Opus 2014/13/B/ST2/02543, 2014/15/B/ST2/03998, and 2015/19/B/ST2/02861, Sonata-bis 2012/07/E/ST2/01406; the National Priorities Research Program by Qatar National Research Fund; the Programa Severo Ochoa del Principado de Asturias; the Thalis and Aristeia programmes cofinanced by EU-ESF and the Greek NSRF; the Rachadapisek Sompot Fund for Postdoctoral Fellowship, Chulalongkorn University and the Chulalongkorn Academic into Its 2nd Century Project Advancement Project (Thailand); the Welch Foundation, contract C-1845; and the Weston Havens Foundation (USA).
\end{acknowledgments}

\bibliography{auto_generated}

\cleardoublepage \appendix\section{The CMS Collaboration \label{app:collab}}\begin{sloppypar}\hyphenpenalty=5000\widowpenalty=500\clubpenalty=5000\textbf{Yerevan Physics Institute,  Yerevan,  Armenia}\\*[0pt]
A.M.~Sirunyan, A.~Tumasyan
\vskip\cmsinstskip
\textbf{Institut f\"{u}r Hochenergiephysik,  Wien,  Austria}\\*[0pt]
W.~Adam, F.~Ambrogi, E.~Asilar, T.~Bergauer, J.~Brandstetter, E.~Brondolin, M.~Dragicevic, J.~Er\"{o}, M.~Flechl, M.~Friedl, R.~Fr\"{u}hwirth\cmsAuthorMark{1}, V.M.~Ghete, J.~Grossmann, J.~Hrubec, M.~Jeitler\cmsAuthorMark{1}, A.~K\"{o}nig, N.~Krammer, I.~Kr\"{a}tschmer, D.~Liko, T.~Madlener, I.~Mikulec, E.~Pree, N.~Rad, H.~Rohringer, J.~Schieck\cmsAuthorMark{1}, R.~Sch\"{o}fbeck, M.~Spanring, D.~Spitzbart, W.~Waltenberger, J.~Wittmann, C.-E.~Wulz\cmsAuthorMark{1}, M.~Zarucki
% ...
\vskip\cmsinstskip
% ...
1:~~Also at Vienna University of Technology, Vienna, Austria\\
% ...

\end{sloppypar}
\end{document}